\documentclass{sig-alternate2}
\pdfoutput=1
\usepackage[charter]{mathdesign}
\usepackage[mathcal]{eucal}

\usepackage{hyperref}
\usepackage{tikz}
\usetikzlibrary{arrows,shapes,decorations,automata,calc}

\usepackage{xspace} 
\usepackage{bbold} 
\usepackage{ntheorem}
\usepackage{graphicx}
\usepackage[labelfont=bf]{caption}
\usepackage{subcaption}
\usepackage{url}

\pgfdeclarelayer{edgelayer}
\pgfdeclarelayer{nodelayer}
\pgfsetlayers{edgelayer,nodelayer,main}

\tikzstyle{none}=[inner sep=0pt]

\tikzstyle{rn}=[circle,fill=Red,draw=Black,line width=0.8 pt]
\tikzstyle{gn}=[circle,fill=Lime,draw=Black,line width=0.8 pt]
\tikzstyle{yn}=[circle,fill=Yellow,draw=Black,line width=0.8 pt]

\tikzstyle{simple}=[-,line width=2.000]
\tikzstyle{arrow}=[-,postaction={decorate},decoration={markings,mark=at position .5 with {\arrow{>}}},line width=2.000]
\tikzstyle{tick}=[-,postaction={decorate},decoration={markings,mark=at position .5 with {\draw (0,-0.1) -- (0,0.1);}},line width=2.000]
\tikzstyle{edge}=[->,draw]
\tikzstyle{thin}=[-,draw]
\tikzstyle{width}=[<->,draw]

\renewcommand{\exp}[1]{\ensuremath{\mathrm{e}^{#1}}}

\definecolor{amber}{rgb}{1.0, 0.49, 0.0}

\definecolor{arsenic}{rgb}{0.23, 0.27, 0.29}

\definecolor{azure}{rgb}{0.0, 0.5, 1.0}

\DeclareMathOperator{\dblone}{\ensuremath\mathbb{1}}
\newtheorem{theorem}{Theorem} 
\newtheorem{lemma}{Lemma}
\newtheorem{definition}{Definition}

\newcounter{examp}
\newtheorem{example}[examp]{Example}

\newcommand\Reals{\ensuremath{\mathbb{R}}}
\newcommand\Realspo{\ensuremath{\mathbb{R}_{\geq 0}}}

\newcommand\Nats{\ensuremath{\mathbb{N}}}

\newcommand{\Exp}[1]{\ensuremath{ \mathrm{e}^{#1} }}

\renewcommand{\det}{\textrm{\textbf{det}}\xspace}

\newcommand{\kibam}{\textsc{KiBaM}\xspace}
\newcommand{\kib}{\textsc{\textbf{K}}\xspace}
\newcommand{\kibfull}{\bar{\kib}\xspace}
\newcommand{\kibbounds}{\kib^\Box\xspace}
\newcommand{\kibapprox}{\kib^{\approx\Box}\xspace}

\newcommand{\bfull}{\bar{b}\xspace}
\newcommand{\soc}{\textsc{SoC}\xspace}

\newcommand{\x}{a}
\newcommand{\y}{b}
\newcommand{\X}{A}
\newcommand{\Y}{B}

\newcommand{\amax}{\ensuremath{a_{\max}}}
\newcommand{\bmax}{\ensuremath{b_{\max}}}
\newcommand{\btresh}[1]{\ensuremath{b_{\mathrm{tresh}}(#1)}}
\newcommand{\Itop}{\ensuremath{\bar{I}}\xspace}

\newcommand{\densSoC}[1]{\ensuremath{f_{#1}}}
\newcommand{\densFull}[1]{\ensuremath{\bar{f}_{#1}}}
\newcommand{\dead}[1]{\ensuremath{z_{#1}}}
\newcommand{\densLoad}{g}

\newcommand{\mtp}{\ensuremath{\mathcal{M}}}
\newcommand{\durations}{\ensuremath{\Delta}}
\newcommand{\densities}{\ensuremath{\mathbf{g}}}

\newcommand{\ca}[1]{\ensuremath{\mathrm{#1}_{a}}}
\newcommand{\cb}[1]{\ensuremath{\mathrm{#1}_{b}}}
\newcommand{\cax}{\ensuremath{\ca{q}}}
\newcommand{\cay}{\ensuremath{\ca{r}}}
\newcommand{\cai}{\ensuremath{\ca{s}}}
\newcommand{\cbx}{\ensuremath{\cb{q}}}
\newcommand{\cby}{\ensuremath{\cb{r}}}
\newcommand{\cbi}{\ensuremath{\cb{s}}}

\newcommand{\e}{\ensuremath{\mathrm{e}^{-kt}}}

\newcommand{\capacity}{d}

\newcommand{\probm}{\mathrm{\mathbf{Pr}}}

\begin{document}
%

\title{
Recharging Probably Keeps Batteries Alive%
}
%
%
%
%
%

\numberofauthors{3} 
%
\author{
%
%
%
\alignauthor
Holger Hermanns\\
       \affaddr{Saarland University}\\
       \affaddr{Saarbr\"ucken, Germany}
\alignauthor
Jan Kr\v{c}\'{a}l\\
       \affaddr{Saarland University}\\
       \affaddr{Saarbr\"ucken, Germany}
\alignauthor
Gilles Nies\\
       \affaddr{Saarland University}\\
       \affaddr{Saarbr\"ucken, Germany}
}

\sloppy 

\maketitle
\abstract{
  The \emph{kinetic battery model} is a popular model of the dynamic
  behavior of a conventional battery, useful to predict or optimize
  the time until battery depletion. The model however lacks certain
  obvious aspects of batteries in-the-wild, especially with respect to
  $(\mathit{i})$ the effects of random influences and $(\mathit{ii})$ the behavior when
  charging up to capacity bounds.

  This paper considers the kinetic battery model with bounded capacity
  in the context of piecewise constant yet random charging and
  discharging.  The resulting model enables the time-dependent
  evaluation of the risk of battery depletion. This is exemplified in
  a power dependability study of a nano satellite mission.
}




\section{Introduction}
The \emph{kinetic battery model} (\kibam) is a popular representation
of the dynamic behavior of the \emph{state-of-charge} (\soc) of a
conventional rechargeable battery~\cite{manwell1993lead,rao}. Given a
constant load, it characterizes the battery \soc by two coupled
differential equations. Empirical evaluations show that this model
provides a good approximation of the \soc across various battery
types~\cite{so75079,DBLP:journals/iee/JongerdenH09}.

The original \kibam does not take capacity bounds into considerations,
it can thus be interpreted as assuming infinite capacity. Reality is
unfortunately different.
When studying the \kibam operating with capacity bounds, it becomes
apparent that charging and discharging are \emph{not} dual to each
other. However, 
opposite to the discharging process, the charging
process near capacity bounds has not received dedicated attention in
the literature. That problem is attacked in the present paper.

Furthermore, statistical results obtained by experimenting with real
of-the-shelf batteries suggest considerable variances in actual
performance~\cite{buchmann}, likely rooted in manufacturing 
and wear differences. This observation asks for a stochastic
re-interpretation of the classical \kibam to take the statistically
observed \soc spread into account on the model level, and this is what
the present paper develops -- in a setting with capacity bounds. 
It views the \kibam as a transformer of the
continuous probability distribution describing the \soc at any real time point, thereby also supporting uncertainty and noise in the load process.

The approach presented not only enables the treatment of randomness
with respect to the battery itself, but also makes it possible to
determine the \soc distribution after a sequence of piecewise
constant, yet \emph{random} charge or discharge loads. We develop the
approach in a setting with continuous randomness so as to directly
support normal (i.e.\ Gaussian), Weibull or exponential distributions.  We
apply the model to a case study inspired by a nano satellite currently
orbiting the earth~\cite{gomspacewebsite}, for which we need to
superpose it with a periodic deterministic charge load, representing
the infeed from on-board solar panels.

The resulting battery model can be viewed as a particular stochastic hybrid
system \cite{abate2008probabilistic,altman1997asymptotic,blom2006stochastic,bujorianu2005bisimulation,davis1984piecewise,sproston2000decidable}, developed without discretizing time. It
can (for instance) for any given real time point provide probabilistic
guarantees about the battery never being depleted before.

The genuine contributions of the paper are: $(\mathit{i})$ The interpretation
of the \kibam as a transformer of \soc distributions, $(\mathit{ii})$ developed
without discretizing time, $(\mathit{iii})$ considering both charging and
discharging in the context of capacity bounds, $(\mathit{iv})$ applied to a
case study of a low earth orbiting satellite.

\paragraph*{Related work}
Haverkort and Jongerden~\cite{DBLP:journals/iee/JongerdenH09}
review broad research on various battery models. They  
discuss \emph{stochastic} battery
models~\cite{rao,DBLP:conf/dsn/ClothJH07} which view the \kibam for a
given load as a stochastic process, unlike our (more accurate) view as
a deterministic transformer of the randomized initial conditions of
the battery.
Furthermore, in this survey, the problem of charging bounds does not get dedicated attention. 


Battery capacity has been addressed only by Boker \emph{et
  al.}~\cite{boker2014battery}. They considered a discretized,
unbounded \kibam together with a possibly non-deterministic and cyclic
load process, synthesizing initial capacity bounds to power the process safely. Hence,
capacity is here understood as an over-dimensioned initial condition and not
as a truly limiting charging bound.

\emph{Random loads} on a battery, generated by a continuous-time Markov chain, have been previously studied by Cloth
\emph{et al.}~\cite{DBLP:conf/dsn/ClothJH07}. Their setting cannot be easily extended by charging since they view the available and bound charge levels as two types of \emph{accumulated reward} in a reward-inhomogeneous
continuous time Markov chain. 
%

An extension of the \kibam to \emph{scheduling} has been considered by
Jongerden \emph{et al.} \cite{jongerden2009maximizing}. They compute
optimal schedules for multiple batteries in a discretized setting with
only discharging. This has been taken up and improved using techniques
from the planning domain \cite{DBLP:conf/ijcai/FoxLM11}.
\section{The Kinetic Battery Model}
\begin{figure}
\centering
\includegraphics[width=0.45\textwidth]{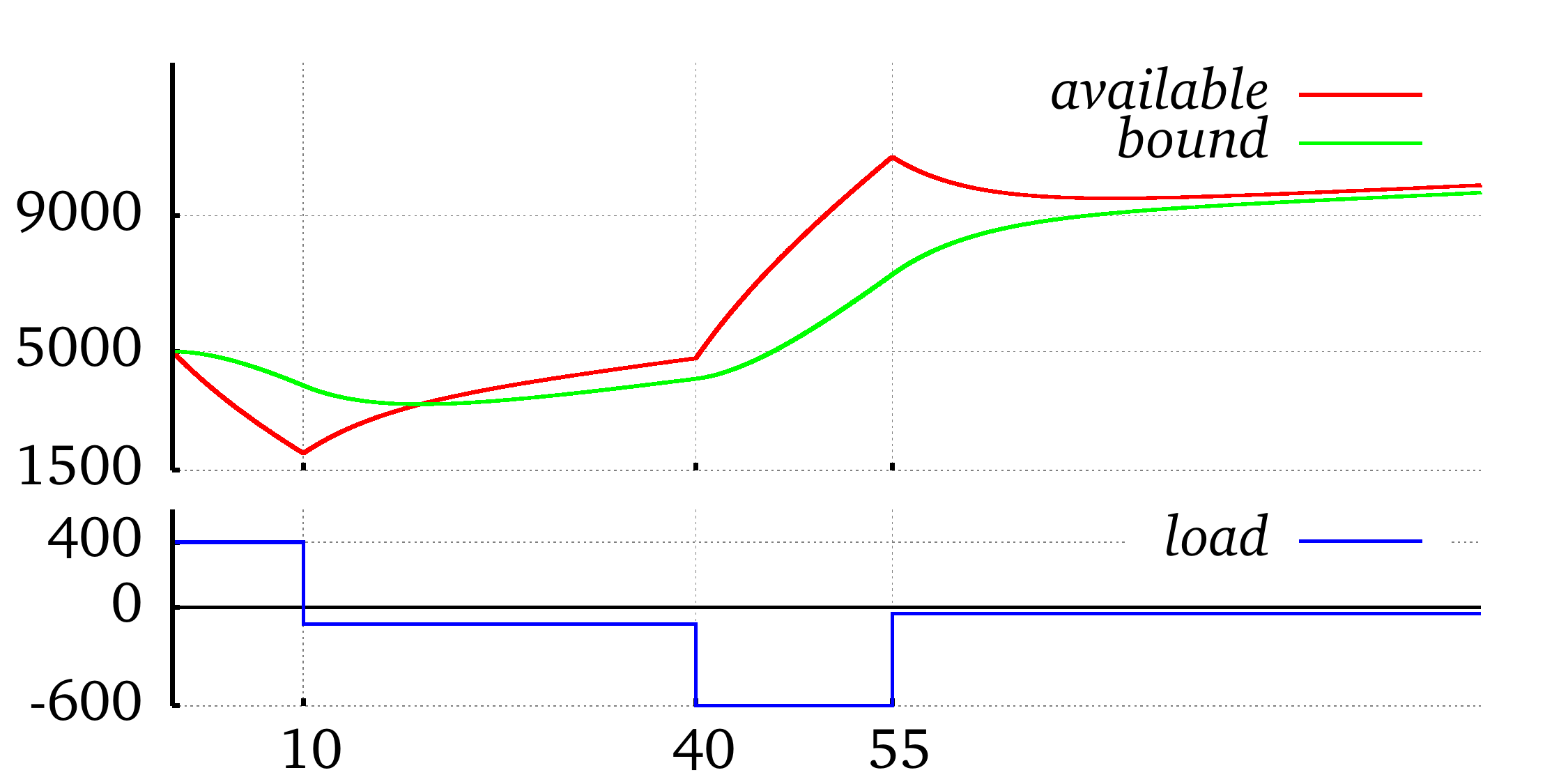}
\caption{Evolution of the state of charge as time passes (top) with
  the battery strained by a piecewise-constant load (bottom). The
  initially available charge decreases heavily due to the load $400$
  but the restricted diffusion makes the bound charge
  decrease only slowly up to time $10$; after that the battery
  undergoes a mild recharge, followed by a strong recharge and a mild 
  recharge at the end. At all times the bound charge approaches the available
  charge by a speed proportional to the difference of the two values.}
\label{Fig:simple-kibam-plot}
\end{figure}

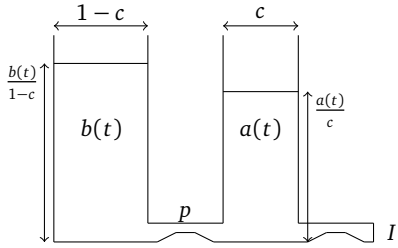
\begin{figure}
\centering
\begin{tikzpicture}[xscale=0.5,yscale=0.5]
\begin{pgfonlayer}{nodelayer}
		\node [style=none] (0) at (-3, 3.25) {};
		\node [style=none] (1) at (-0.5, 3.25) {};
		\node [style=none] (2) at (1.5, 3.25) {};
		\node [style=none] (3) at (3.5, 3.25) {};
		\node [style=none] (4) at (3.5, -1.75) {};
		\node [style=none] (5) at (5.5, -1.75) {};
		\node [style=none] (6) at (5.5, -2.25) {};
		\node [style=none] (7) at (5.25, -2.25) {};
		\node [style=none] (8) at (3.75, -2.25) {};
		\node [style=none] (9) at (4.75, -2) {};
		\node [style=none] (10) at (4.25, -2) {};
		\node [style=none] (11) at (1.25, -2.25) {};
		\node [style=none] (12) at (0.75, -2) {};
		\node [style=none] (13) at (0.25, -2) {};
		\node [style=none] (14) at (-0.25, -2.25) {};
		\node [style=none] (15) at (-3, -2.25) {};
		\node [style=none] (16) at (-0.5, -1.75) {};
		\node [style=none] (17) at (1.5, -1.75) {};
		\node [style=none] (18) at (-3.25, 2.5) {};
		\node [style=none] (19) at (-3.25, -2.25) {};
		\node [style=none] (20) at (3.75, -2.25) {};
		\node [style=none] (21) at (3.75, 1.75) {};
		\node [style=none] (22) at (-3, 2.5) {};
		\node [style=none] (23) at (-0.5, 2.5) {};
		\node [style=none] (24) at (1.5, 1.75) {};
		\node [style=none] (25) at (3.5, 1.75) {};
		\node [style=none] (26) at (0.5, -1.5) {$p$};
		\node [style=none] (27) at (-3.85, 2) {$\frac{b(t)}{1-c}$};
		\node [style=none] (28) at (4.35, 1.25) {$\frac{a(t)}{c}$};
		\node [style=none] (29) at (6, -2) {$I$};
		\node [style=none] (30) at (-3, 3.5) {};
		\node [style=none] (31) at (-0.5, 3.5) {};
		\node [style=none] (32) at (1.5, 3.5) {};
		\node [style=none] (33) at (3.5, 3.5) {};
		\node [style=none] (34) at (-1.75, 3.85) {$1-c$};
		\node [style=none] (35) at (2.5, 3.85) {$c$};
		\node [style=none] (36) at (-1.75, 0.75) {$b(t)$};
		\node [style=none] (37) at (2.5, 0.75) {$a(t)$};
	\end{pgfonlayer}
	\begin{pgfonlayer}{edgelayer}
		\draw [style=thin] (3.center) to (4.center);
		\draw [style=thin] (4.center) to (5.center);
		\draw [style=thin] (5.center) to (6.center);
		\draw [style=thin] (6.center) to (7.center);
		\draw [style=thin] (7.center) to (9.center);
		\draw [style=thin] (9.center) to (10.center);
		\draw [style=thin] (10.center) to (8.center);
		\draw [style=thin] (8.center) to (11.center);
		\draw [style=thin] (11.center) to (12.center);
		\draw [style=thin] (12.center) to (13.center);
		\draw [style=thin] (13.center) to (14.center);
		\draw [style=thin] (14.center) to (15.center);
		\draw [style=thin] (15.center) to (0.center);
		\draw [style=thin] (1.center) to (16.center);
		\draw [style=thin] (16.center) to (17.center);
		\draw [style=thin] (17.center) to (2.center);
		\draw [style=thin] (22.center) to (23.center);
		\draw [style=thin] (24.center) to (25.center);
		\draw [style=width] (19.center) to (18.center);
		\draw [style=width] (20.center) to (21.center);
		\draw [style=width] (30.center) to (31.center);
		\draw [style=width] (32.center) to (33.center);
	\end{pgfonlayer}
\end{tikzpicture}
\caption{The two-well-model of the \kibam with the available charge on the right (exposed directly to the load $I$) connected to the bound charge on the left by a pipe of width~$p$.}
\label{Fig:TwoWellsModel}
\end{figure}

The kinetic battery model is a mathematical characterization of the
state of charge of a battery. It differs from an ideal energy source
by incorporating the fact that not all the energy stored in a battery
is available at all times. The stored energy is divided into two
portions, the \emph{available charge} and the \emph{bound charge}.
Only the available charge may be consumed immediately by a
\emph{load} supported by the battery and thereby behaves similar to an
idealized source. As time passes, some of the bound charge is
converted into available charge and is thus free to be consumed. This
effect is coined the \emph{recovery effect} as the available charge
\emph{recovers} to some extend during periods of low discharge or no
discharge at all. The recovery effect agrees with our experiences
using batteries: For instance, when a cellphone switches off due to an
apparently empty battery, it often can be switched back on after
waiting a few minutes. The battery seems to have recovered. This
\emph{diffusion} between available and bound charge can take place in
either direction depending on the amount of both types of energy
stored in the battery. Thus, while charging the battery, available
charge is converted to bound charge. This behavior is illustrated by Figure~\ref{Fig:simple-kibam-plot}.
\paragraph*{Coupled differential equations}
The \kibam is often depicted as two wells holding liquid, the
available charge and the bound charge well, interconnected by a pipe
that represents the diffusion of the two types of charge, see Figure
\ref{Fig:TwoWellsModel}.
Formally, the \kibam is characterized by two coupled differential equations.
\begin{eqnarray}
\dot{a}(t) &=&  -I + p\left(\frac{b(t)}{1-c} - \frac{a(t)}{c}\right) \label{availDE1}\\
\dot{b}(t) &=&  p\left(\frac{a(t)}{c}-\frac{b(t)}{1-c}\right)\label{boundDE1}
\end{eqnarray}
Here, the functions $a(t)$ and $b(t)$ describe the \textit{available}
and \textit{bound} charge respectively, $I$ is a \emph{load}
on the battery, $p$ is the diffusion rate between both wells and $c$
is the width of the available charge well. Thus $1-c$ is the width of
the bound charge well. Intuitively $a(t)/c$ and
$b(t)/(1-c)$ are the level of the fluid stored in the available
charge well and the bound charge well, respectively.  By defining \[
k= \dfrac{p}{c(1-c)}\] we can rewrite (\ref{availDE1}) and
(\ref{boundDE1}) to
\begin{eqnarray*}
\dot{a}(t) &=&  -I + c k \cdot b(t) - (1-c)  k \cdot  a(t) \label{availDE2}\\
\dot{b}(t) &=&  (1-c)k\cdot a(t)-ck \cdot b(t). \label{boundDE2}
\end{eqnarray*}
We will use this version of the \kibam~ODEs throughout this paper.

\paragraph*{Solving the equations}
Using Laplace transforms the \kibam ODE system can be solved, arriving
at \begin{eqnarray*}
a_{t,I}(\x_0,\y_0) &= &\cax(t) \x_0 + \cay(t) \y_0 + \cai(t) I \label{availSol}\\
b_{t,I}(\x_0,\y_0) &= & \cbx(t) \x_0 + \cby(t) \y_0 + \cbi(t) I \label{boundSol}
\end{eqnarray*}
where $\x_0$ and $\y_0$ are the initial available and bound charge levels and the time-dependent coefficients of $\x_0, \y_0$ and $I$ in the equations can be expressed as
\begin{eqnarray*}
\cax(t) &=& \phantom{-}(1-c)\e + c\\
\cbx(t) &=& -(1-c)\e + (1-c) \\
\cay(t) &=& -c\cdot\e \hspace{16pt} + c\\
\cby(t) &=& \phantom{-}c\cdot\e \hspace{16pt} + (1-c) \\
\cai(t) &=& \frac{(1-c)(\e-1)}{k} - t \cdot c \\
\cbi(t) &=& \frac{(1-c)(1-\e)}{k} - t \cdot (1-c).
\end{eqnarray*}
From the solution we can see that the \kibam is affine in $\x_0$ and $\y_0$ (and also $I$). Thus we can combine the two functions into one vector
valued linear mapping
\begin{equation*}
\kib_{t,I}\left[ \begin{array}{c}
\x_0 \\ 
\y_0
\end{array} \right] = \left[ \begin{array}{ccc}
\cax(t) & \cay(t) & \cai(t) \\ 
\cbx(t) & \cby(t) & \cbi(t)
\end{array} \right] \cdot \left[ \begin{array}{c}
\x_0 \\ 
\y_0 \\ 
I
\end{array} \right].
\label{KibamVec}
\end{equation*}
When $t$ is clear from context, we simplify the notation and drop 
the argument of these coefficients (without dropping the time-dependency).
From now on we prefer semicolon notation $[\x;\y]$ to denote column vectors. (All vectors appearing in this paper are column vectors.) Furthermore, whenever we compare two vectors, e.g., $[\x;\y] \leq [\x',\y']$, we interpret the order component-wise.

\begin{example}\label{ex:1}
  The function $\kib$ can be used to approximate the final \soc in
  Figure~\ref{Fig:simple-kibam-plot} (for $k = 1/100$, $c = 1/2$, and
  $\circ$ denoting function composition) by
\begin{align*}
\left[\x;\y\right] &= \kib_{44,100} \circ \kib_{15,-600} \circ \kib_{30,-100}\circ \kib_{10,400} \left[ 5000; 5000 \right] \\
&\approx \kib_{44,100} \circ \kib_{15,-600} \circ \kib_{30,-100} \left[ 2002; 3998 \right] \\
&\approx \kib_{44,100} \circ \kib_{15,-600} \left[ 4802; 4198 \right] \\
&\approx \kib_{44,100} \left[ 10732; 7268 \right], \\
\intertext{and with the last step in more details (denoting $\exp{-\frac{44}{100}}$ by $E$),}
&= \left[ \begin{array}{ccc}
\phantom{-}\frac{1}{2}E + \frac{1}{2} & - \frac{1}{2}E + \frac{1}{2} & 50E-50 - \frac{44}{2} \\ 
-\frac{1}{2}E + \frac{1}{2} \rule{0pt}{2.6ex} & \phantom{-} \frac{1}{2}E + \frac{1}{2} & 50-50E - \frac{44}{2}
\end{array} \right] \cdot \left[ \begin{array}{c}
10732 \\ 
7268 \\ 
-35
\end{array} \right] \\
&
=
\left[ \begin{array}{c}
-18 E + 6480 \\ 
\phantom{-}18 E + 8020
\end{array} \right]
\approx 
\left[ \begin{array}{c}
9881 \\ 
9659 
\end{array} \right].
\end{align*}
On the last line, 
the first summands (with $E$) stand for the spread of the values when the recovery effect has not converged yet (as for $t\to\infty$, $E\to 0$). For $c=1/2$ and zero load, the recovery effect makes the difference of the available charge and the bound charge converge to $0$. However, for non-zero load $I$, it does not converge to $0$ but to $I/k$ which explains the difference in the second summands.
\end{example}

\paragraph*{Battery Depletion}

A standard application of \kibam is to find out whether a task can be
performed with a given initial state of charge without depleting the
battery. A task is a pair $(T,I)$ with $T$ being the task execution
time, and $I$ representing the load, imposed for duration $T$.

For an execution time $T$ and a load $I$, we say that a battery with a \soc $[\x_0;\y_0] > [0;0]$ \emph{powers a task} $(T,I)$ if 
\begin{align*}
\kib_{t,I} \left[\x_0;\y_0\right] & > [0;0] & \text{$\forall \; 0 \leq t \leq T$}.
\end{align*}

Let us stress that the state of charge of the battery evolves in negative numbers in the same way as in positive numbers because the differential equations do not have any explicit bounds. 
To rule out that the \soc of the battery goes into negative numbers and returns back, we need a certain form of monotonicity.


%

\setcounter{examp}{0} 
\begin{example}[Cont.]
We notice that neither the available nor the bound charge are monotonous in the standard sense. In Figure~\ref{Fig:simple-kibam-plot}, the bound charge is not monotonous on $[10,40]$, the available charge is not monotonous on $[55,100]$. However, for instance, on $[40,55]$, available charge is the first to get above the value $9000$ (and never crosses the boundary again).
\end{example}

\begin{lemma}\label{lem:monotonous}
For any $\kappa,I\in\Reals$, $\preceq \; \in \{\leq,\geq\}$, and $0 \leq t \leq T$,
\begin{align*}
\forall \x_0 \not\preceq \kappa, \forall \y_0 \not\preceq \kappa: \;\; \y_{t,I}(\x_0, \y_0) \preceq \kappa \; &\Longrightarrow \; \x_{t,I}(\x_0, \y_0) \preceq \kappa, \\
\forall \x_0 \not\preceq \kappa, \forall \y_0 \not\preceq \kappa: \;\; \x_{t,I}(\x_0, \y_0) \preceq \kappa \; &\Longrightarrow \; \x_{T,I}(\x_0, \y_0) \preceq \kappa.
\end{align*}
\end{lemma}
Intuitively speaking, the first property states that the available charge is always the first to cross a bound, the second property states that when the available charge crosses a bound it never returns back (for a given load). 

As a direct consequence of Lemma~\ref{lem:monotonous}, we can easily figure out whether the battery powers a task.

\begin{lemma}
A battery with a \soc $[\x_0;\y_0] > [0;0]$ powers a task $(T,I)$ if and only if $\kib_{T,I} \left[\x_0;\y_0\right] > [0;0]$.
\end{lemma}

\section{The Basic Random KiBaM}
Some basic notions from probability theory are needed for the further development. 
Let $f_X$ and $f_{X \times Y}$ denote the density function of a random variable $X$ and the joint density function of a pair of random variables $(X,Y)$, respectively.
%

The \emph{conditional density function} $f_{X \mid Y}$ of $X$ given the occurrence of the value $y$ of $Y$
%
is defined as
$f_{X \mid Y} (x|y) = f_{X \times Y}(x,y)/f_Y(y)$. From this expression for $f_{X \times Y}$, we obtain by marginalization the density function $f_X$ as
%
\begin{equation}
f_X(x) = \int_{-\infty}^{\infty} f_{X \mid Y} (x|y) f_Y(y) ~ \mathrm{d}y.
\label{Def:marginaldensity_with_conditional}
\end{equation}


Furthermore, the \emph{transformation law for random variables} enables the
construction of unknown density functions from known ones given the relation between the corresponding random variables. Formally,
for every $d$-dimensional random vector $\mathbf{X}$ and every
injective and continuously differentiable function $g:\Reals^d \to
\Reals^d$, we can express the density function of $\mathbf{Y} :=
g(\mathbf{X})$ at value $y$ in the range of $g$ as
\begin{equation}
f_{\mathbf{Y}}(y) = f_{\mathbf{X}}\left(g^{-1}(y)\right) \cdot \left| \det \left(J_{g^{-1}}(y)\right) \right|
\label{Def:TransformationLaw}
\end{equation}
where $J_f(x)$ denotes the \emph{Jacobian} of a mapping $f$ evaluated
at $x$. Let us recall that the Jacobian of $f$ is the matrix of the partial derivatives of the mapping $f$.

\paragraph*{Joint Density of the State of Charge}
In order to consider the \kibam as a stochastic object, it appears
natural to consider the vector $[\x_0;\y_0;I]$ as being random. This
naturally reflects the situation where the initial state of the
battery is subject to perturbations due to manufacturing or wear
variances, and so is its load. Therefore, we assume the initial \soc is expressed by random variables
$\X_0, \Y_0$ endowed with a joint probability density function $\densSoC{0}$ and that the load on the battery is expressed by a random variable $I$ endowed with a probability density function $\densLoad$. We assume that the random variables $I$
and $(\X_0,\Y_0)$ are independent. 

\begin{example} \label{exa:two}
A second running example addresses the random \kibam. Instead of a
single (Dirac) \soc, we now assume that the joint density $\densSoC{0}$
of the charge is, say, uniform over the area $[4,6.5]\times[4,6.5]$ (below).
\begin{center}
\includegraphics[width=0.55\linewidth]{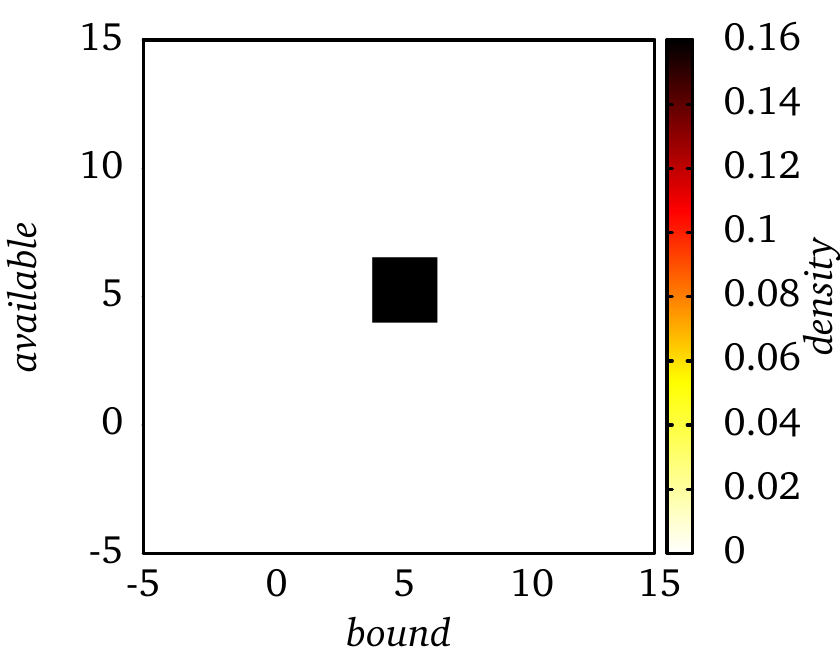}
\end{center}
We shall illustrate our findings how the \soc distribution
evolves as the time passes on this particular example.
\end{example}
Let $(\X_T,\Y_T)$ denote the random variables expressing \soc after time $T$ of \emph{constant} (but random) load $I$. 
We are interested in the joint probability distribution of $(\X_T,\Y_T)$ Thus, for a
given time point $T$ we want to establish the joint density function
of the vector $[\X_T;\Y_T]$ given by
\begin{equation}
\left[ \begin{array}{c}
\X_T \\ 
\Y_T \\ 
\end{array} \right] = \left[ \begin{array}{ccc}
\cax & \cay & \cai \\ 
\cbx & \cby & \cbi
\end{array} \right] \cdot \left[ \begin{array}{c}
\X_0 \\ 
\Y_0 \\ 
I
\end{array} \right].
\label{KibamVec_t}
\end{equation}
Expressing the joint density using direct application of the
transformation law for random variables is not possible because the mapping is not invertible.
%
However, using the fact that $I$ and
$(\X_0,\Y_0)$ are independent, 
it is still possible to use the transformation
law of random variables so that ultimately we arrive at an analytic
characterization of the joint density of $(\X_T,\Y_T)$.

In the following we will derive the conditional density of $(\X_T,\Y_T)$
under the condition that $I=i$ for some arbitrary but fixed value
$i$. As $\densLoad$ is known, we afterwards accommodate for the missing
information about $I$ via integration over the range of $I$. 
Knowing that $I=i$ eliminates one source of randomness, (\ref{KibamVec_t}) can be rewritten
to
\begin{equation*}
\left[ \begin{array}{c}
\X_T \\ 
\Y_T \\ 
\end{array} \right] = \left[ \begin{array}{cc}
\cax & \cay \\ 
\cbx & \cby
\end{array} \right] \cdot \left[ \begin{array}{c}
\X_0 \\ 
\Y_0
\end{array}\right] +
\left[ \begin{array}{c}
\cai \\ 
\cbi
\end{array} \right] \cdot
i
\label{KibamVec_conditional_on_i}
\end{equation*}
which is an invertible linear mapping and thus allows to express the joint density of $(A_T,B_T)$ in terms of the density of  $(A_0,B_0)$ via the transformation law of random variables. 

Inverting this mapping $\kib_{T,i}$ results in 
\begin{align*}
&\kib_{T,i}^{-1} \left[ \begin{array}{c}
\x \\ 
\y \\ 
\end{array} \right] = 
\exp{kT}
\left[
\begin{array}{ccc}
 \cby & -\cay & \cay\cbi-\cby\cai \\ 
 -\cbx & \cax & \cbx\cai-\cax\cbi
\end{array}
\right] \cdot \left[ \begin{array}{c}
\x \\ 
\y \\
i
\end{array} \right].
\end{align*}
By a straightforward computation, the determinant of the Jacobian of $\kib_{T,i}^{-1}$ is $\det{J_{\kib_{T,i}^{-1}}} = \exp{kT}$.
Note that it is constant in $a$, $b$, and $i$, it only depends on $T$. 
%
%
Thus, using (\ref{Def:TransformationLaw}) we arrive at the joint density
of $(A_T,B_T)$ conditioned by $I=i$ 
\begin{align*}
\densSoC{T}(\x,\y\mid i) 
= 
& f_{\kib_{T,i} [\X_0;\Y_0]}(\x,\y) 
\\ 
= & 
\densSoC{0}\left(\kib_{T,i}^{-1}[\x;\y]\right)
\cdot \left| \exp{kt} \right| 
\end{align*}
where $f_{\kib_{T,i} [\X_0;\Y_0]}$ denotes the joint density of the random vector $\kib_{T,i} [\X_0;\Y_0]$.
According to (\ref{Def:marginaldensity_with_conditional}) we
arrive at the unconditional density over $(\X_T,\Y_T)$ via integration on $i$:
\begin{lemma}\label{lem:prob-kibam}
Let $T$ be execution time and $\densLoad$ be load density.
For an initial \soc $\densSoC{0}$ over $(\X_0,\Y_0)$ and task $(T,\densLoad)$, the joint distribution of $(\X_T,\Y_T)$ is absolutely continuous with density $\densSoC{T}$ given by
\begin{align*}
\densSoC{T} (x,y)
& = \int_{-\infty}^{\infty} \densSoC{0} \left(\kib_{T,i}^{-1}[x;y]\right)
\cdot \exp{kT} \cdot \densLoad(i) ~ \mathrm{d}i.
\end{align*}
\end{lemma}

\setcounter{examp}{1} 
\begin{example}[Cont.]
  We return to our example assuming the density $\densLoad$
  of the load being uniform between $[-0.1,0.1]$.  Based on the
  expression from Lemma~\ref{lem:prob-kibam}, we can compute the \soc
  of the battery after task $(20,\densLoad)$, displayed on the left,
  and $(60,\densLoad)$, displayed on the right. We arbitrarily chose the parameters $c=0.5$ and $p=0.002$.\\
                \includegraphics[width=0.49\linewidth]{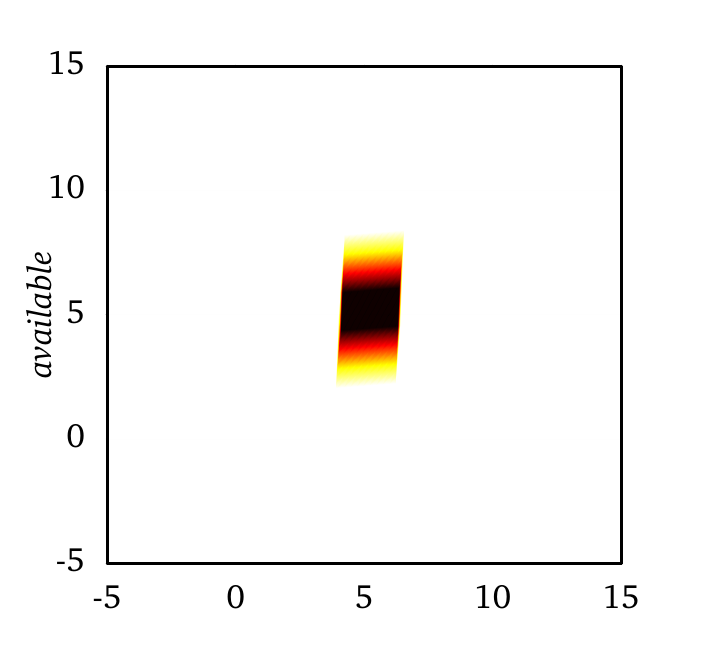}~
                \includegraphics[width=0.49\linewidth]{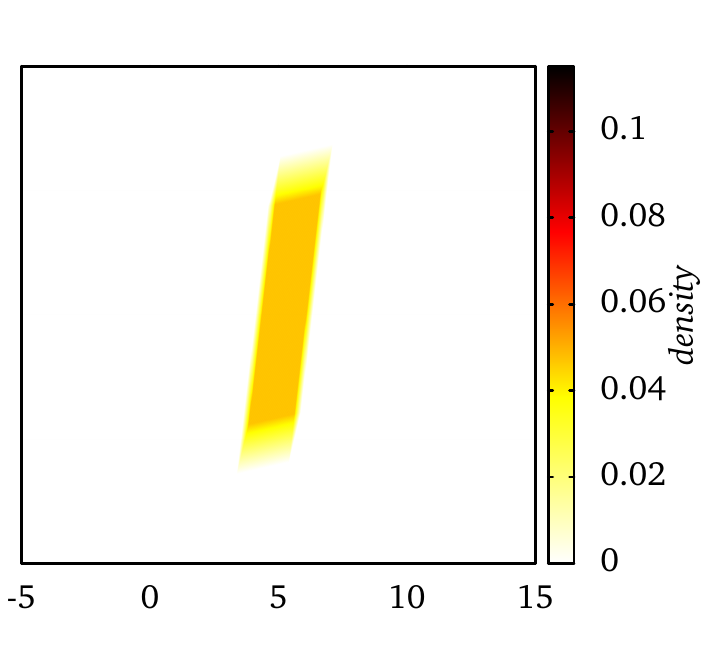}
\end{example}

\paragraph*{Risk of Depletion}
Let us transfer the problem of battery depletion into the stochastic setting.
We say that a density $\densSoC{0}$ is \emph{positive} if for any $\x,\y$ such that either $\x \leq 0$ or $\y \leq 0$ we have $\densSoC{0}(\x,\y) = 0$.
For an execution time $T > 0$ and a load density $\densLoad$, we say that the battery with positive \soc $\densSoC{0}$ \emph{powers with probability $p>0$} a task $(T,\densLoad)$ if
\begin{align*} 
\probm \left[ \forall 0 \leq t \leq T: (\X_t,\Y_t) > (0,0) \right] \; \geq \; p.
\end{align*}
%

Due to the monotonicity of \kibam from Lemma~\ref{lem:monotonous}, this is equivalent to observing the probability of being empty \emph{only} at time $T$.
From Lemma~\ref{lem:prob-kibam} we obtain the following.
\begin{lemma}
\label{lem:lemm}
A battery with \soc $\densSoC{0}$ powers with probability $p>0$ a task $(T,\densLoad)$ if and only if
\begin{align*}
 \int_{0}^{\infty} \int_{0}^{\infty}
  \densSoC{T} (\x,\y) ~ \mathrm{d}\y ~ \mathrm{d}\x
  \; \geq \;
  p.
\end{align*}
\end{lemma}

\setcounter{examp}{1} 
\begin{example}[Cont.]
  Thanks to the lemma, it
  suffices to perform the integration on the densities displayed in the
  previous plots in this running example. The probability to power the
  tasks $(20,\densLoad)$ is $1$, for the task $(60,\densLoad)$ it is just
$\approx 0.968$.
\end{example}

%
\section{Bounded Recharging} 
For the evaluation of the long-run state of a battery, a good
understanding of the charging process is as essential as understanding the
discharging process. Both are well supported by the theory developed
so far, and have occurred in our examples in the form of negative
loads.  What is not treated in the theory yet is a capacity bound of
the battery which is a real constraint in most applications.
To the best of our knowledge, charging in \kibam while respecting its capacity restrictions has not been addressed even in the deterministic case. 
This is what we are going
to develop first, and then extend to the randomized setting.

Let us assume that the battery has capacity $\capacity$ divided into capacity $\amax = c\cdot\capacity$ of the available charge well and capacity $\bmax = (1-c)\cdot\capacity$ of the bound charge well.

\paragraph*{Charging at full available charge}

A battery with empty available charge can no longer support its
task. Notably, a battery with full available charge does not behave
dual to the behavior at empty available charge, just because a
battery with full available charge \emph{continues to operate}, and we
thus need to consider its further charging.  In this discussion we do
not consider crystallization effects reported for some batteries when
overcharging~\cite{buchmann,PhysRevLett.81.4660}. These phenomena do
affect battery wear, not considered in our modeling efforts thus far.

When the available charge reaches its capacity $\amax = c\cdot \capacity$ and
is still charged further by (high enough) incoming current, its value
stays constant and only the bound charge increases due to
diffusion. Hence, we know that $\dot{a}(t) =
0$ and $\frac{a(t)}{c} = \capacity$ and we can thus modify (\ref{boundDE1})
to

\begin{align}
\dot{b}(t) &= p\left(\capacity-\frac{b(t)}{1-c}\right) 
 \label{boundDE3}
 \tag{2'}.
\end{align}

Note that this equation describes the behavior of the battery at time $t$ only  if the load satisfies 
$$
- I \geq \dot{b}(t).
$$
Since $I$ is constant and the diffusion is decreasing over time, the inequality holds for all $t$, provided 
$$ -I \geq \dot{b}(0) = p\left(\capacity-\frac{b(0)}{1-c}\right),$$
or equivalently if $b(0) \geq  \btresh{I}$ where
\begin{align}
\btresh{I} = \bmax + I\cdot \frac{1-c}{p}. \label{minimalBoundCharge}
\end{align}
For a smaller initial bound charge, the standard differential equations (\ref{availDE1}) and (\ref{boundDE1}) apply, until the available charge hits the capacity bound again.
Finally, solving the ODE (\ref{boundDE3}) yields

\begin{lemma}\label{lem:charging-with-full}
Let $T$ be an execution time, $I$ be a load, and $\y_0$ be a bound charge such that 
$\y_0 \geq \btresh{I}$. A battery with a \soc $(\amax,\y_0)$ reaches after the task $(T,I)$ the state of charge $(\amax, \bfull_T(\y_0))$ where
\begin{align}
\label{Def:bound_full}
\bfull_T(\y_0) = \Exp{-ckT}\y_0 + \left( 1- \Exp{-ckT} \right) \cdot \bmax
\end{align}
and $k$ again stands for $p/\left(c\cdot(1-c)\right)$. 
\end{lemma}

We notice that the resulting bound charge $\bfull_T(\y_0)$ does not further depend on $I$, i.e. one cannot make the battery charge faster by increasing the charging current.
Furthermore, for a fixed $\y_0$, the curve of $t \mapsto \bfull_t(\y_0)$ is a negative exponential starting from the point $\y_0$ with the full capacity $\bmax$ of the bound charge being its limit. Thus, Lemma~\ref{lem:charging-with-full} also reveals that the bound charge in finite time never gets full and there is no need to describe this situation separately.
Finally, we denote analogously by 
 $\kibfull_T\left[\x_0;\y_0\right] = \left[\x_0; \bfull_T(\y_0) \right]$ the linear mapping describing the behavior at the upper bound.

\setcounter{examp}{0} 
\begin{example}[Cont.]
If we put an upper bound of $9000$ to the battery scenario from Figure~\ref{Fig:simple-kibam-plot}, the battery ends up with a slightly smaller charge at time $100$.

\includegraphics[width=0.95\linewidth]{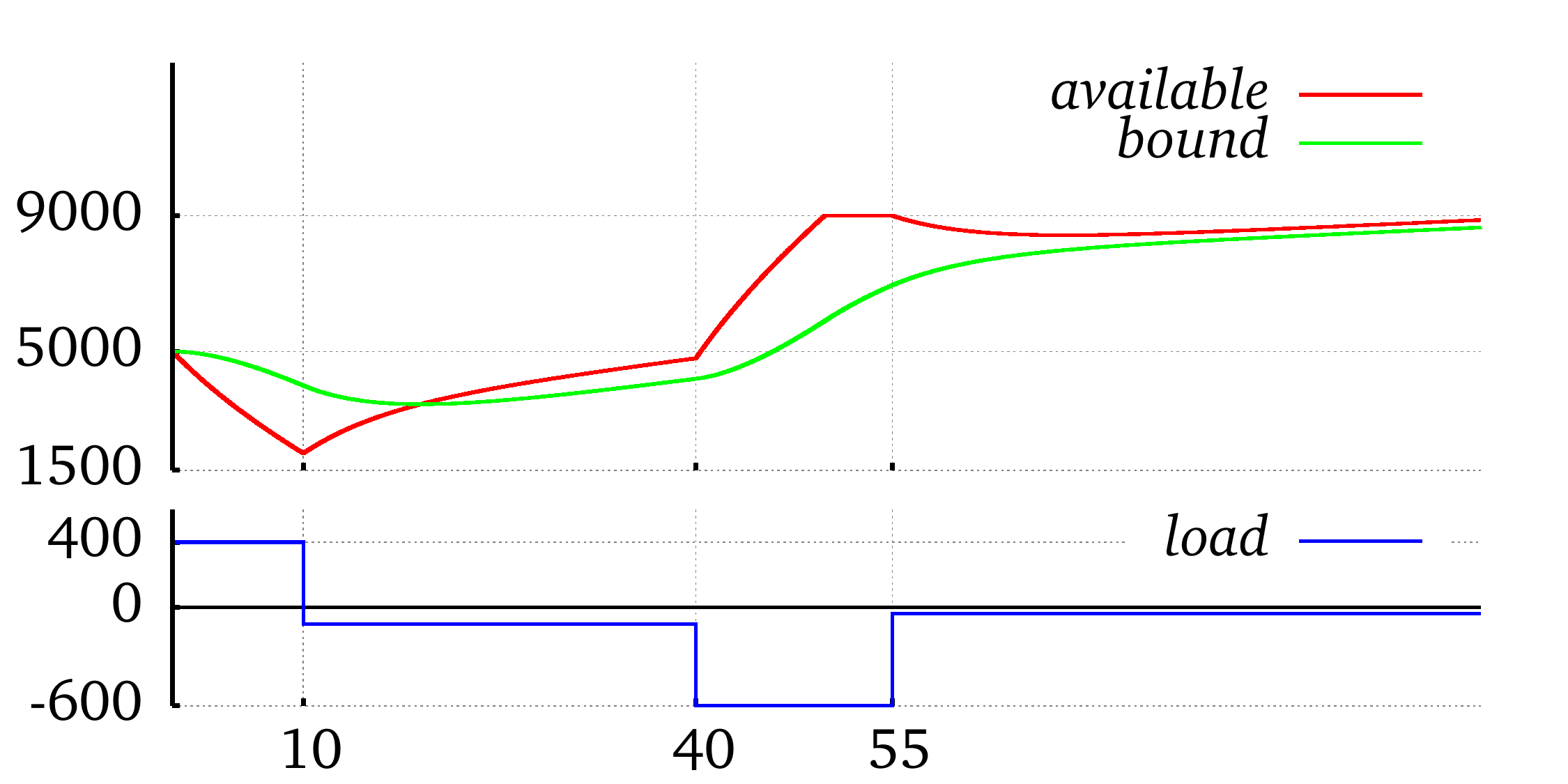}
\end{example}

\paragraph*{Hitting the capacity bound}

For a given constant load $I$, we have seen two types of behavior of the
battery: ($i$) \emph{before} it hits the available charge capacity and ($\mathit{ii}$)
\emph{after} it hits the capacity. The remaining question is
\emph{when} it hits the capacity limit. For a given initial state $(\x_0,\y_0)$ and a load $I$, this amounts to solving
\begin{align*}
a_{t,I}(\x_0,\y_0) = \amax,
\end{align*}
which in turn yields an equation
\begin{align*}
u \cdot \exp{-kt} + v \cdot t + w &= 0
\end{align*}
where 
$u = \x_0\left(1-c\right) -\y_0c + (c+1) \cdot I/k$, $v = -Ic$, and $w = cd-\x_0c -\y_0c - (1-c)\cdot I/k$. This can be solved as 
\begin{align}\label{eq:hitting}
t = -W\left(\frac{u}{v}\cdot \exp{-\frac{w}{v}} \right) - \frac{w}{v}
\end{align}
where $W$ is the product log function. It has no closed form but can be arbitrarily numerically approximated\cite{DBLP:journals/adcm/CorlessGHJK96}.


\paragraph*{Deterministic \kibam with lower and upper bounds}

All the previous building blocks allow us to express easily the \soc
of a deterministic \kibam after powering a given task $(T,I)$ when considering battery bounds.
We define it as the following function:
\begin{align*}
\kibbounds_{T,I} \left[ \x_0;\y_0 \right]  =  \begin{cases}
\kib_{T,I} \left[\x_0;\y_0\right]
& \text{if $\x_0 > 0$ and } \\
& \text{$\;\;0 < \x_{T,I}(\x_0,\y_0) \leq \amax$,} \\
\kibfull_{t} \circ \kib_{\bar{t},I} \left[ \x_0;\y_0\right]
& \text{if $\x_0 > 0$ and } \\
& \text{$\;\;\x_{T,I}(\x_0,\y_0) > \amax$,} \\
\left[0;0\right]
& \text{otherwise, i.e. if $\x_0 = 0$} \\
& \text{\;\;or $\x_{T,I}(\x_0,\y_0) < 0$}
\end{cases}
\end{align*}
where $\bar{t}$ is the largest solution of (\ref{eq:hitting}) and $t = T-\bar{t}$.

The \kibam evolution has one fundamental property we rely on heavily later: for any fixed time and load, it is monotonous with respect to starting \soc.
\begin{lemma}
Let $(\x,\y)$ and $(\x',\y')$ be two \soc{s}.
For every $t>0$ and for every $I \in \Reals$ it holds that
$$(\x,\y) \leq (\x',\y') ~ \Longrightarrow ~ \kibbounds_{t,I}[\x;\y] \leq \kibbounds_{t,I}[\x';\y'].$$
\label{lem:preservingorder}
\end{lemma}
%

\setcounter{examp}{0} 
\begin{example}[Cont.]
By introducing the bounds in the first running example, the computation of the final \soc changes only in the interval $[40,55]$. Here, instead of $\kib_{15,-6}$, we apply $\kib_{\bar{t},-6}$ for the first $\bar{t}\approx 7.8$ time units, followed by $\kibfull_{15-\bar{t}}$.
\end{example}

The computation of the time point $\bar{t}$ is problematic, as mentioned before. An alternative to numerical approximation of the exact time point of crossing a bound is provided by the following observation: If time point $t$ is fixed, we can check whether the available charge will exceed $\amax$ after $t$ time units. In this case it is possible to determine the charging current $\Itop$ necessary for the available charge to hit $\amax$ \emph{exactly} after $t$ time units, i.e. solving
$a_{t,I}(\x_0,\y_0) = \amax$ for $I$ instead of $t$. Let us denote the solution of this equation by $\Itop(\x_0,\y_0)$ for a \soc $(\x_0,\y_0)$ and conclude that it can be computed by
$$
\Itop(\x_0,\y_0) = - \frac\cax\cai \cdot \x_0 - \frac\cay\cai \cdot \y_0 + \frac\amax\cai.
$$
Such a slower charging rate $\Itop(\x_0,\y_0)$ allows us 
to define a conservative under-approximation of the exact \kibam 
$\kibbounds_{T,I}[\x_0;\y_0]$. 
We replace the case $\kibfull_{t} \circ \kib_{\bar{t},I} \left[ \x_0;\y_0\right]$ where the upper bound is reached by
$\kib_{T,\Itop(\x_0,\y_0)}[\x_0;\y_0]$. 
We will henceforth refer to this under-approximation by $\kibapprox_{T,I}$. 
\begin{lemma} For any \soc $(\x,\y)$ we have
$ \kibapprox_{T,I} [\x; \y] \leq \kibbounds_{T,I}[\x;\y].$
\label{lem:underapproximation}
\end{lemma}
Together with Lemma \ref{lem:preservingorder}, the above lemma tells us that, from the moment we used $\kibapprox$ first, we will never overshoot the exact behavior of the \kibam with bounds $\kibbounds$. This fact is illustrated by the next example.

\setcounter{examp}{0} 
\begin{example}[Cont.]
We keep the upper bound of $9000$ but instead of using the exact \kibam behavior 
with bounds $\kibbounds$, we use the $\kibapprox$ approximation. The load in the interval $[40,55]$ is approximately $-432.49$ instead of $-600$ as the available charge would cross its bound. Instead it reaches $9000$ exactly at $t=55$. Note that from here on (i.e. in the interval $(55,100]$) the \soc is not exceeding the corresponding \soc from the previous figure.

\includegraphics[width=0.95\linewidth]{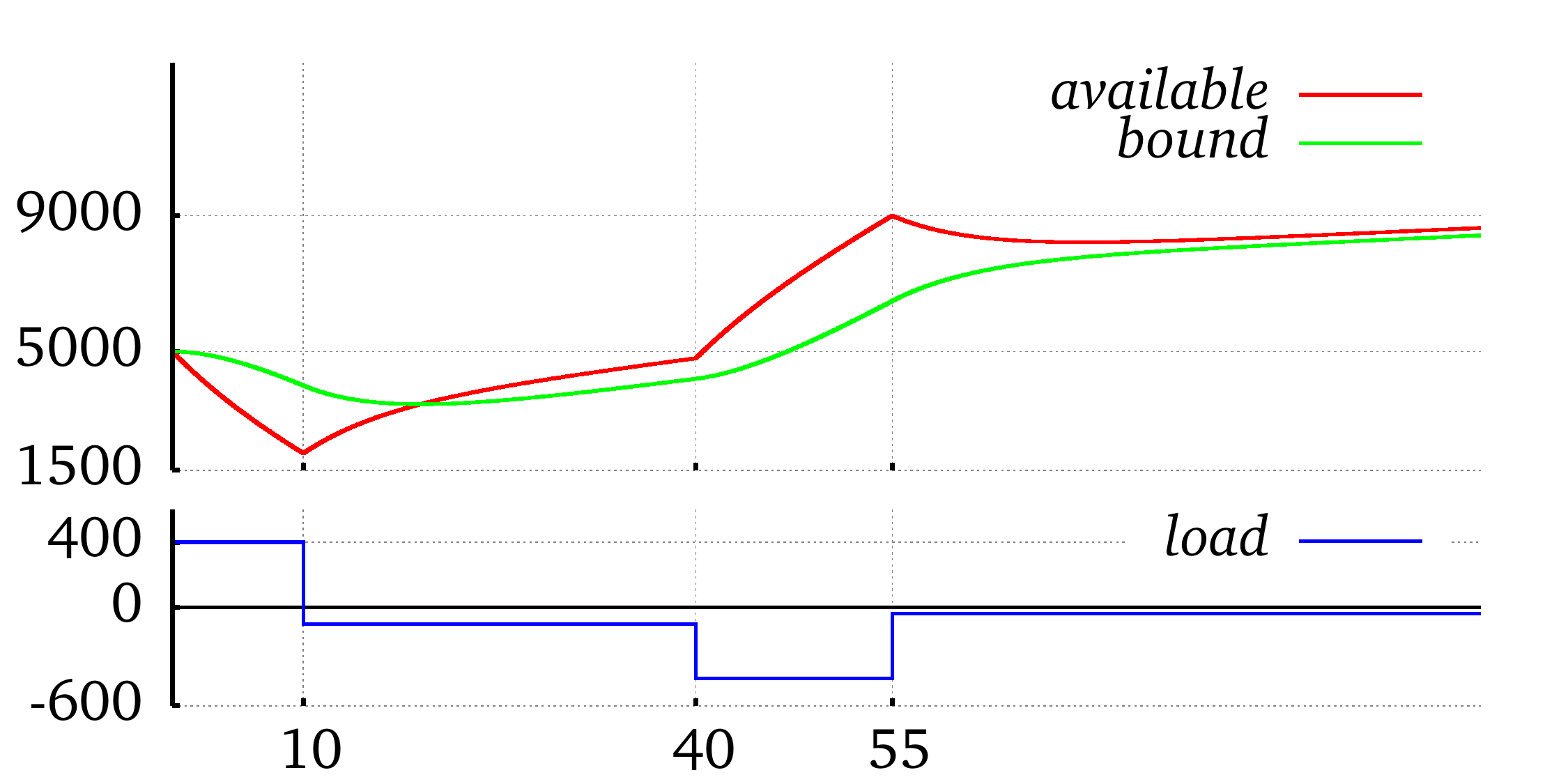}
\end{example}

\paragraph*{Random \kibam with lower and upper bounds}

We now turn our attention to the challenge of 
assuming that the random variables $(A_t,B_t)$ evolve according to 
$\kibbounds_{T,I}$.
We first observe that the 
joint distribution of $(\X_T,\Y_T)$ may not be absolutely continuous, because positive probability may accumulate in the point $(0,0)$ where the battery is empty and on the line
$\{(\amax,\y) \mid 0 < \y < \bmax\}$ where the available
charge is full. Hence, we represent each $(\X_t,\Y_t)$ by a triple $\langle \densSoC{t}, \densFull{t}, \dead{t} \rangle$ where
\begin{itemize}
\item $\densSoC{t}$ is the joint density describing the distribution in the ``inner'' area $(0,\amax) \times (0,\bmax)$,
\item $\densFull{t}$ is the density over bound charge describing the distribution on the upper line $\{\amax\} \times (0,\bmax)$, and
\item $\dead{t} \in [0,1]$ is the probability of being empty
\end{itemize} 
such that for any measurable $A \subseteq \Reals \times \Reals$ we have
\begin{align*}
\probm\left[(\X_t,\Y_t) \in A \right] &= 
  \int_{0}^{\amax} \int_{0}^{\bmax} \densSoC{t}(\x,\y) \dblone_{(\x,\y)\in A} ~ \mathrm{d}\y ~ \mathrm{d}\x \\
 & \; +  \int_{0}^{\bmax} \densFull{t}(\y) \dblone_{(\amax,\y)\in A} ~ \mathrm{d}\y
 + \dead{t} \dblone_{(0,0)\in A}
\end{align*}
where $\dblone_{\varphi}$ denotes the indicator function of a condition $\varphi$.

For an initial \soc $\langle \densSoC{0}, \densFull{0}, \dead{0} \rangle$ and for a given task $(T,\densLoad)$, our aim is to express the resulting \soc $\langle \densSoC{T}, \densFull{T}, \dead{T} \rangle$. 

First, we address the evolution inside the bounds by defining $\densSoC{T}$. 
The new part is to describe how the battery moves from the upper bound to the area inside the bounds. 
We define a vector valued function 
$[b;i] \mapsto \kib_{T,i}\left[ \amax;\y \right]$
that maps (for a \soc with full available charge $\amax$) a pair of bound charge and load $[\y;i]$ to a \soc $[\x;\y]$ in the usual \kibam fashion. 
We denote the inverse of this mapping component-wise as $[\x;\y] \mapsto \left[ \mathcal{B}(\x,\y); \mathcal{I}(\x,\y) \right]$ where the functions $\mathcal{B}(\x,\y)$ and $\mathcal{I}(\x,\y)$ can be explicitly expressed as 
\begin{align*}
\mathcal{I}(\x,\y) &= (\amax \e - \cby \x - \cbx \y) / \left(\cay\cbi-\cby\cai \right),\\
\mathcal{B}(\x,\y) &= -\cbx \x + \cax \y + (\cbx\cai-\cax\cbi) \cdot \mathcal{I}(\x,\y).
\end{align*}
The Jacobian determinant of this inverse map is easily derived to be $1/\left( \cay \cbi - \cai \cby \right)$ and is constant in the \soc and the load.

The joint density has for any $a < \amax$ and $b < \bmax$ the form
\begin{align}\label{eq:kibam-down-from-bounds}
\densSoC{T}(\x,\y) 
\; = \;
& \int_{-\infty}^{\infty} \densSoC{0}\left(\kib^{-1}_{T,i}\left[ \x;\y \right] \right) \cdot | \exp{kT} | \cdot \densLoad(i) ~ \mathrm{d}i  \\
& + \; \densFull{0}(\mathcal{B}(\x,\y)) \cdot |1/\left( \cay \cbi - \cai \cby \right)| \cdot \densLoad(\mathcal{I}(\x,\y)). \notag
\end{align}

\begin{figure}
\begin{tikzpicture}[inner sep = 0,outer sep=0,xscale=0.75,yscale=0.75]

\coordinate (topleft) at (0, 0) {};
\coordinate (topright) at (10, 0) {};
\node [rotate=5](diagonal) at (6.9,-3.1) 
{};

\node(bottomleft) at (0, -2.8) {$\vdots$};
\node (bottomright) at (10, -2.8) {$\vdots$};

\draw [] (topleft) to node [above=15,pos=0.5] {bound charge} (topright);
\path [] (bottomleft) to (bottomright);
\draw [] (bottomleft) to node [above=6, sloped,pos=0.3] {available charge} (topleft);
\draw [] (bottomright) to (topright);
\draw [] (diagonal) to node [below=2, sloped,pos=0.7] {$a=b$} (topright);


\begin{scope}[xshift=17em]

\node [font=\Large, outer sep=.5, circle, label=90:{$(\amax,b)$}] (t1) at (3.2,0) {\textbullet};
\node [font=\Large, label=90:{$(\amax,\bar{b})$}] (s1) at (.8,0) {\textbullet};
\node [font=\Large,label=90:{$(\amax,b')$}] (s7) at (-0.9,0) {\textbullet};
\node [font=\Large,] (s2) at (1.7,-2.5) {\textbullet};

\draw [ultra thick, ->] (s1) to node [above=3,font=\small] {$\bar{\kib}_T$} (t1);
\draw [ultra thick, ->] (s2) to[in=230,out=80] 
	node[font=\Large,pos=0.10] {\textbullet} 
	node[font=\Large,pos=0.27] {\textbullet} 
	node[font=\Large,pos=0.44] {\textbullet} 
	node[font=\Large,pos=0.61] {\textbullet} 
	node[font=\Large,pos=0.78] {\textbullet} 
	node [above=4, sloped,pos=0.5,font=\small] {$\kib_{t,i}$} 
	node [below=8, sloped,pos=0.5,font=\scriptsize] {$\forall 0 \leq t \leq T$} 
(t1);

\node [outer sep=5] (hit) at (1.8, 0.3) {};

\draw [very thin,->,font=\small] (s7) to [out=330,in=210]
 node [below=2,pos=0.4] {$\kib_{T,i}$} 
 (hit);

\end{scope}


\begin{scope}[xshift=-15em]

\node [font=\Large, circle, outer sep=2,label=270:{$\quad\;\;\;(a,b)$}] (t2) at (8.5,-.8) {\textbullet};

\node [font=\Large] (s3) at (6.1,-1) {\textbullet};
\node [below=8] at (6.2,-1) {$(a_i,b_i)$};
\node [font=\Large] (s4) at (6.4,0) {\textbullet};
\node [above=3] at (6.4,0) {$(\amax,\mathcal{B}(a,b))\qquad\qquad\quad\quad$};
\node [circle,draw, thick, inner sep=1.5] (s5) at (7.1,1) {};
\node [circle,draw, thick, inner sep=1.5] (s6) at (6.7,0.55) {};
\coordinate (further) at (6.38, -0.2) {};

\draw [ultra thick, ->] (s3) to[in=210,out=330] 
	 node [below=3] {$\kib_{T,i}$} 
 (t2);

\draw [ultra thick, ->,font=\small] (s4) to[in=190,out=300] 
	 node [sloped,below=-1,pos=0.3,rotate=10] {$\kib_{T,\mathcal{I}(a,b)}$} 
 (t2);

\draw [very thin,->,font=\small] (s5) to 
	 node [sloped,above,pos=0.4] {$\kib_{T,0}$} 
 (t2);

\draw [very thin,->,font=\small] (s6) to [out=305,in=170]
	 node [sloped,above,pos=0.2] {$\kib_{T,1}$} 
 (t2);

\draw [densely dotted] (s5)  to[in=55,out=220] (s6)  to[in=70,out=235] (s4) to (further);

\end{scope}

\end{tikzpicture}
\caption{Illustration of the upper bound in random \kibam.
On the left, we depict the evolution towards a point $(a,b)$ in the inner area of the state space. 
The figure shows different points from which the battery in the given time $T$ reaches (for different loads) the point $(a,b)$. Only one point corresponding to a unique load has available charge $\amax$.
On the right, we show three ways to reach the upper bound.
}
\label{figUpperBounds}
\end{figure}
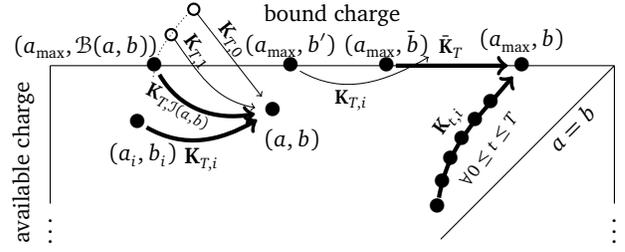

The first summand in (\ref{eq:kibam-down-from-bounds}) comes from the density $\densSoC{0}$ of the inner area by the standard unbounded \kibam. Ranging over all loads $i$, it integrates the density $\densSoC{0}$ of such points $(a_i,b_i)$ that satisfy $\kib_{T,i}[a_i;b_i] = [a;b]$, i.e. $[a_i;b_i] = \kib^{-1}_{T,i}[a;b]$.
Lemma~\ref{lem:monotonous} again guarantees correctness that the bounds are not crossed in the meantime.
The second summand comes from $\bar{f}_0$, due to discharging the battery down from the capacity bound as discussed above. 
%
Both summands are illustrated on the left hand side of Figure~\ref{figUpperBounds}.

After obtaining the density $\densSoC{T}$, we turn to $\densFull{T}$, which is difficult to characterize, we are not aware of any closed-form expression achieving this. Hence we resort at this point to a conservative under-approximation of the battery charge.
For a random \soc $(\X_0,\Y_0)$ given by $\langle \densSoC{0},\densFull{0}, \dead{0}\rangle$ and a task $(T,\densLoad)$ we define $(\X'_T,\Y'_T)$ given by $\langle \densSoC{T}',\densFull{T}', \dead{T}'\rangle$. We define the densities for $(0,0) < (\x,\y) < (\amax, \bmax)$ by
\begin{align*}
\densSoC{T}'(\x,\y) \; = \; 
& \densSoC{T}(\x,\y) \\
\bar{f}'_T(\y) 
\; = \;
& \int_{-\infty}^{\infty} 
\Big[\;
\bar{f}_0 \left( \bar{\y} \right) \cdot \exp{ckT}
\cdot 
\dblone_{\bar{\y} \geq \btresh{i}} \\
& \quad + \densFull{0}(\y) \cdot \dblone_{\y < \btresh{i}} \cdot \dblone_{\x_{T,i}(\amax, \y) > \amax} \\
& \quad + \int_0^T f_0\left(\kib^{-1}_{t,i}\left[ \amax ;\y \right]\right) \cdot \exp{kt} ~\mathrm{d}t 
\;\Big] \cdot \densLoad(i) ~\mathrm{d}i \\
\dead{T}' \; = \;
& \int_{-\infty}^{0} \int_{-\infty}^{0} \densSoC{T}(\x,\y) ~\mathrm{d}\y ~\mathrm{d}\x 
\end{align*}
where $\bar{\y}$ denotes $\bfull^{-1}_T(\y) = \exp{ckT}\cdot \y + (1-\exp{ckT})(1-c)d$.

For expressing $\densSoC{T}'$ and $\dead{T}'$, we use the exact
evolution given by (\ref{eq:kibam-down-from-bounds}). 
Let us now closer discuss the density $\densFull{T}'$ at the upper bound which is an integral over all loads $i$.


The first summand in the integral comes from the density $\densFull{0}$ of a point $(\amax,\bar{b})$ at the capacity bound
such that $\kibfull_T[\amax;\bar{b}] = [\amax;b]$. This summand is taken into account only for such
$(\amax,\bar{b})$ where the charging current covers the diffusion so
that the battery evolves along the capacity bound as expressed by
Lemma~\ref{lem:charging-with-full}. 
Technically, we again apply the transformation law for random variables.

Let us now address the case that the diffusion in a state $(\amax,b')$ at the upper bound is stronger than the charging current so that the available charge sinks in the beginning but before time $T$ it again hits the upper capacity in some state $(\amax,b'')$. We are not able to express $b'$ using $b''$; hence, we cannot ``move'' the density from $b'$ to $b''$. As  apparent in the second summand, we thus underapproximate here the bound charge by assuming that the density \emph{stays} in such state $(\amax,b')$.

The third summand in $\densFull{T}'$ comes from the density $f_0$ of the inner area and is another under-approximation of bound charge. If available charge of the battery reaches the capacity bound \emph{before} time $T$, we assume that \emph{until} time $T$ the \soc does not further evolve. 
In particular, the density goes to state $(\amax,b)$ from all states $(\x_0,\y_0)$ such that 
%
$$\kib_{t,i} \left[\x_0;\y_0\right] = \left[\amax;\y\right] \quad\text{for some $0\leq t \leq T$.}$$
%
All three summands are illustrated in Figure~\ref{figUpperBounds} on the right. Let us finally state that $(\X'_T,\Y'_T)$ is an under-approximation of $(\X_T,\Y_T)$.

\begin{lemma}\label{lem:random-kibam-with-bounds}
 For a random \soc $(\X_0,\Y_0)$ given by $\langle \densSoC{0},\densFull{0}, \dead{0}\rangle$, a task $(T,\densLoad)$ defining $(\X_T,\Y_T)$ and $(\X'_T,\Y'_T)$, and any fixed \soc $(\x,\y)$,
%
$$\probm([\X_T;\Y_T] \geq [\x,\y]) \;\; \geq \;\; \probm([\X'_T;\Y'_T] \geq [\x,\y]).$$
\end{lemma}
\setcounter{examp}{1} 
\begin{example}[Cont.]
Based on 
Lemma~\ref{lem:random-kibam-with-bounds}, we can approximate the \soc of the random battery from our second running example for 
\emph{battery bounds} $[0,10]$. We consider the same tasks, 
$(20,\densLoad)$ on the left and $(60,\densLoad)$ on the right.\\[0.5em]
\includegraphics[width=0.4905\linewidth]{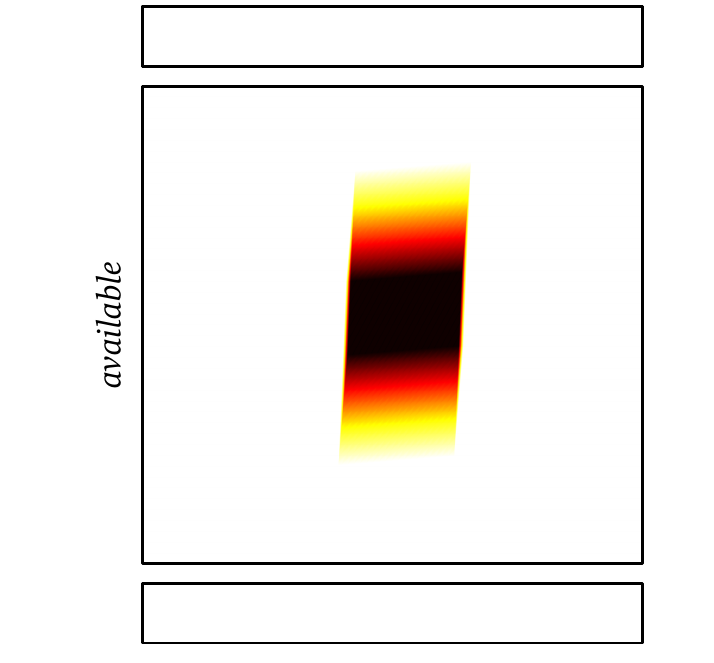}
\includegraphics[width=0.4905\linewidth]{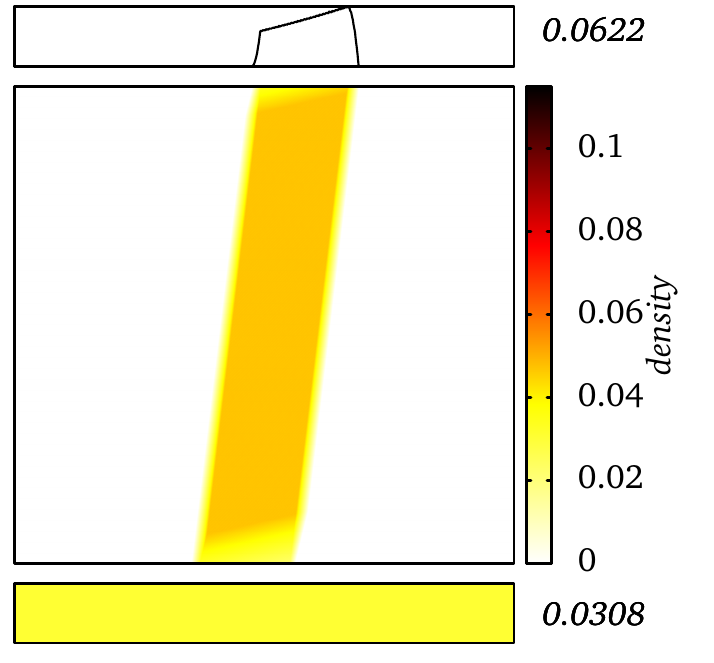}\\[0.5em]
\noindent
The bounded area of the joint density $\densSoC{T}$ is depicted by the largest box. In the small box above we display the density $\densFull{T}$ at the capacity limit. The numbers above and below are the probabilities of available charge being full and empty, respectively (the color below corresponds to the probability).
\end{example}

\section{Markov Task Process}\label{sec:mtp}

So far, we have only discussed execution of one task with fixed duration and random load. In this section, we give a discrete-time Markov model that randomly generates tasks that we call \emph{Markov task process} (MTP).
The formalism is closely inspired by stochastic task graph
models~\cite{sahner1987performance}  or data-flow formalisms such as SDF \cite{lee1987synchronous} or
SADF \cite{theelen2006scenario}. In SDF, task durations are deterministic, and
thus directly supported in our framework. In SADF, durations are in
general governed by discrete probability distributions, which can be translated into our framework at the price of a larger state space.

\begin{definition}
A \emph{Markov task process (MTP)} is a tuple $\mtp = (S,P,\pi, \durations, \densities)$ where $S$ is a finite set of tasks, $P: S\times S \to [0,1]$ is a probability transition matrix, $\pi$ is an initial probability distribution over $S$, $\durations : S \to \Nats$ assigns to each task an integer time duration, and $\densities$ assigns to each task a probability density function of the load.
\end{definition}

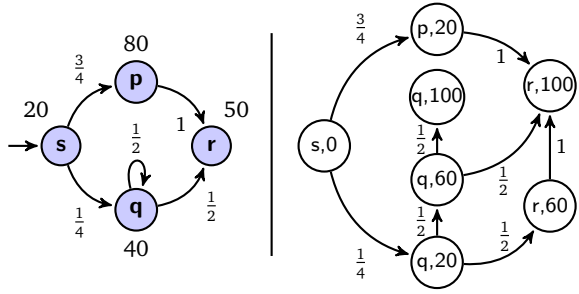
\begin{figure}
\centering

 \begin{tikzpicture}[->,>=stealth',shorten >=1pt,auto,node distance=2.2cm,
  thick,main node/.style={circle, minimum size=1.7em,fill=blue!20,draw,font=\sffamily\bfseries}]

\begin{scope}[xshift=-2.8cm]
  \node[main node,label=100:{$20$}] at (0,0) (1) {s};
  \node[main node,label=90:{$80$}] at (1,0.85) (2) {p};
  \node[main node,label=-90:{$40$}] at (1,-.85) (3) {q};
  \node[main node,label=80:{$50$}] at (2,0) (4)  {r};

\path ($(1)+(-0.7,0)$) edge (1);

  \path[every node/.style={font=\sffamily\small}]
	(1) edge [bend left] node [auto,pos=0.7] {$\frac{3}{4}$} (2)
	(1) edge [bend right] node [auto,pos=0.7,swap] {$\frac{1}{4}$} (3)
	(3) edge [bend right] node [auto,pos=0.7,swap] {$\frac{1}{2}$} (4)
	(2) edge [bend left] node [auto,pos=0.7,swap] {$1$} (4)
	(3) edge [loop above] node [auto] {$\frac{1}{2}$} (3)
;
\end{scope}

\draw[-] (0,1.5) -- (0,-1.5);

\begin{scope}[xshift=0.7cm, every node/.style={draw,circle,font=\sffamily\small,inner sep=0, minimum width=2.25em}]
  \node[] at (0,0) (1) {s,0};
  \node[] at (1.5,1.55) (2) {p,20};
  \node[] at (1.5,-1.55) (3) {q,20};
  \node[] at (3,-0.80) (4)  {r,60};
  \node[] at (1.5,-0.45) (5)  {q,60};
  \node[] at (1.5,0.65) (6)  {q,100};
  \node[] at (3,.80) (7)  {r,100};

  \path[every node/.style={font=\sffamily\small}]
	(1) edge [bend left] node [auto,pos=0.7] {$\frac{3}{4}$} (2)
	(1) edge [bend right] node [auto,pos=0.7,swap] {$\frac{1}{4}$} (3)
	(3) edge [bend right] node [above=-1] {$\frac{1}{2}$} (4)
	(3) edge [] node [auto] {$\frac{1}{2}$} (5)
	(5) edge [] node [auto] {$\frac{1}{2}$} (6)
	(5) edge [bend right] node [below=-1,pos=0.4] {$\frac{1}{2}$} (7)
	(4) edge  node [right=-1] {$1$} (7)
	(2) edge [bend left=20] node [below=-1] {$1$} (7)
;
\end{scope}

\end{tikzpicture}
\caption{An MTP model on the left (with $\durations$ depicted next to states) and its induced graph for $T=100$ on the right.}
\label{fig:mtp}
\end{figure}

An example of a MTP is depicted in Figure~\ref{fig:mtp}.  Intuitively,
a Markov task process $\mtp$ together with an initial distribution
over \soc given by $\langle \densSoC{0},\densFull{0},\dead{0} \rangle$
behaves as follows.  First, an initial \soc $(\x_0,\y_0)$ of the
battery and an initial task $s_0\in S$ are chosen independently at
random according to $\langle \densSoC{0},\densFull{0},\dead{0}
\rangle$, and $\pi$, respectively. Then, the load $i_0$ in task $s_0$
is picked randomly according to $\densities(s_0)$. After the battery
is strained by the load $i_0$ for $\durations(s_0)$ time units, the
process moves into a random successor task $s_1$ (where any $s_1$ is
chosen with probability $P(s_0,s_1)$). Here, the load $i_1$ is
randomly chosen and so on.

Formally, $\mtp$ and $\langle \densSoC{0},\densFull{0},\dead{0} \rangle$ induce a probability measure $\probm$ over samples of the form $\omega = [(\x_0,\y_0);(s_0,i_0)(s_1,i_1)\cdots]$ where the first component is the initial \soc of the battery and the second component describes an infinite execution of $\mtp$. Here, each $s_j$ is the $j$-th task and $i_j$ is the load that is put on the battery for $\durations(s_j)$ time units while performing $s_j$. 
For a given $T\in\Realspo$, the \soc of the battery at time $T$ is expressed by random variables $\X_T,\Y_T$ that are for any $\omega = [(\x_0,\y_0);(s_0,i_0)(s_1,i_1)\cdots]$ defined as
\begin{align*}
\left[
\begin{array}{c}
\X_T(\omega)\\
\Y_T(\omega)
\end{array}
\right]
= \;
&\kibbounds_{t',i_n} \circ \kibbounds_{t_{n-1},i_{n-1}} \circ 
 \;\; \cdots \circ \kibbounds_{t_{0},i_{0}} 
\left[
\begin{array}{c}
\x_0\\
\y_0
\end{array}
\right]
\end{align*}
where each $t_j$ stands for $\durations(s_j)$, and $n$ is the minimal number such that the $n$-th task is not finished before $T$, i.e. $t_n > t'$ where $t' := T - \sum_{j=0}^{n-1}t_j$.

\begin{definition}
We say that a battery with \soc $\langle \densSoC{0}, \densFull{0}, \dead{0}\rangle$ \emph{powers with probability $p>0$} a system $\mtp$ for time $T$ if
\begin{align*} 
\probm \left[ \X_T > 0 \right] \; \geq \; p.
\end{align*}
\end{definition}

In order to (under-)approximate the probability that $\mtp$ is powered for a given time, we need to symbolically express the distribution over $(\X_T,\Y_T)$.
We present an algorithm that builds upon the previous results.

\paragraph*{Expressing the distribution of $(\X_T,\Y_T)$}

Let us fix an \textbf{input} MTP $\mtp = (S,P,\pi, \durations,
\densities)$, distribution over \soc $\langle \densSoC{0},
\densFull{0}, \dead{0}\rangle$, and time $T > 0$.  We consider the
joint distribution of \soc and the MTP.
Intuitively, we split the distribution of \soc into subdistributions and move them along the paths of $\mtp$ according to the probabilistic branching of the MTP. We notice that we do not need to explore all exponentially many paths; when two paths visit the same state at the same moment, we can again merge the two subdistributions.
This process is formalized by the following graph and a procedure how to propagate the distribution through the graph.

For a given MTP $\mtp$ we define a directed acyclic graph $(V,E)$ over
$V = S \times (\{0,1,\ldots,\lfloor T \rfloor \} \cup \{T\})$ such
that there is an edge from a vertex $(s,t)$ to a vertex $(s',t')$ if
$P(s,s')>0$, $t < t'$, and $t' = \min\{t+\durations(s),T\}$. Further,
let $(V',E')$ be the graph obtained from $(V,E)$ by removing vertices
that are not reachable from any $(s,0)$ with $\pi(s) > 0$ (see
Figure~\ref{fig:mtp}).
\begin{enumerate}
\setlength\itemsep{0.1em}
\item We label each vertex of the form $(s,0)$ where $\pi(s) > 0$ by a subdistribution $\langle \densSoC{0}\cdot\pi(s),\densFull{0}\cdot\pi(s),\dead{0}\cdot \pi(s) \rangle$.
\item We repeat the following steps as long as possible.
\begin{enumerate}
\setlength\itemsep{0.1em}
\item For each vertex $(s,t)$ labeled by $\langle \densSoC{},\densFull{},\dead{} \rangle$, we obtain $\langle \densSoC{}',\densFull{}',\dead{}' \rangle$ by Lemma~\ref{lem:random-kibam-with-bounds} for a task $(t'-t,\densities(s))$ where $t' = \min\{ t+\durations(s),T \}$. Then we label each outgoing \emph{edge} from $(s,t)$ to  some $(s',t')$ by $$\left\langle \densSoC{}' \cdot P(s,s'),\;\densFull{}' \cdot P(s,s'),\;\dead{}'\cdot P(s,s') \right\rangle.$$
\item For each vertex $(s,t)$ such that \emph{all} its ingoing edges are labeled by $\langle \densSoC{}^{1},\densFull{}^{1},\dead{}^{1} \rangle, \ldots, \langle \densSoC{}^{k},\densFull{}^{k},\dead{}^{k} \rangle$, we label $(s,t)$ by 
$$\left\langle \densSoC{}^{1} + \cdots + \densSoC{}^{k},\; \densFull{}^{1} + \cdots + \densFull{}^{k}, \;\dead{}^{1} + \cdots + \dead{}^{k} \right\rangle.$$
\end{enumerate}
\end{enumerate}
\noindent
Finally, let all vertices of the form $(s,T)\in V'$ be labeled by $\langle \densSoC{}^{1},\densFull{}^{1},\dead{}^{1} \rangle, \ldots, \langle \densSoC{}^{n},\densFull{}^{n},\dead{}^{n} \rangle$.
The \textbf{output} distribution $\langle \densSoC{T},\densFull{T},\dead{T} \rangle$ is equal to
$$\left\langle \densSoC{}^{1} + \cdots + \densSoC{}^{n},\; \densFull{}^{1} + \cdots + \densFull{}^{n},\; \dead{}^{1} + \cdots + \dead{}^{n} \right\rangle.$$

We naturally arrive at the following theorem.
\begin{theorem}
A battery with \soc $\langle \densSoC{0}, \densFull{0}, \dead{0}\rangle$ powers a system $\mtp$ for time $T$ with probability at least $1-\dead{T}$.
\end{theorem}
\section{The Random KiBaM in Practice}\label{sec:case}
In this section, we apply the results established in the previous
sections in a concrete scenario. The problem is inspired by
experiments currently being carried out with an earth orbiting nano
satellite, the
GOMX-1 \cite{gomspacewebsite}.

\paragraph*{Satellite}
GOMX-1 \cite{gomspacewebsite} is a Danish two-unit CubeSat mission
launched in November 2013 to perform research and experimentation in
space related to Software Defined Radio (SDR) with emphasis on
receiving ADS-B signal from commercial aircraft over oceanic areas. As
a secondary payload the satellite flies a NanoCam C1U color camera
for earth observation experimentation. Five sides are covered with
NanoPower~P110 solar panels, the power system NanoPower~P31u holds a
$7.4$V Li-Ion battery of capacity $5000$~mAh. GOMX-1 uses a radio
amateur frequency for transmitting telemetry data, making it possible
to receive the satellite data with low-cost infrastructure anywhere on
earth.
The mission is developed in collaboration between GomSpace ApS, DSE
Airport Solutions and Aalborg University, financially supported by the
Danish National Advanced Technology Foundation. The empirical studies
carried out with GOMX-1 serve as a source for parameter values and
motivate the scenario described in the sequel.
We concretely use the following data collected from extensive
in-flight telemetry logs.
\begin{itemize}
\item One orbit takes 99 minutes and is nearly polar; 
\item The battery capacity is $\capacity = 5000$ mAh;   
\item During $4$ to $7$ out of on average $15$ orbits per day,
  communication with the base station takes place.  The load induced
  by communication is roughly $400$ mA. The length of the
  communication depends on the distance of the pass of the satellite
  to the base station and varies between $5$ and $15$ minutes;
\item In each communication, the satellite can receive instructions on
  what activities to perform next. This influences the subsequent
  background load. Three levels of background load dominate the logs,
  with average loads at $250$ mA, $190$ mA, and $90$ mA. These
  background loads subsume the power needed for operating the
  respective activities, together with basic tasks such as sending
  beacons every $10$ seconds;
\item Charging happens periodically, and spans around $2/3$rd of the
  orbiting time. Average charge power is $400$~mA;
\end{itemize}
The above empirical observations determine the base line of our
modeling efforts. Still the case study described below is a synthetic
case. We make the following assumptions:
 
\begin{itemize}
%
\item We assume constant battery temperature. The factual temperature
  of the orbiting battery oscillates between -8 and 25 degree Celsius
  on its outside. There is the (currently unused) on-board option to
  heat the battery to nearly constant temperature. Using an on-off
  controller, this would lead to another likely nearly periodic load
  on the battery, well in the scope of what our model supports.
\item A constant charge from the solar panels is assumed when exposed
  to the sun. The factual observed charge slowly decays.
  This is likely caused by the fact that solar panels operate
  better at lower temperature (opposite to batteries), but heat up
  quickly when coming out of eclipse. 
%
\item We assume a strictly periodic charging behavior. The factual
  charging follows a more complicated pattern determined by the
  relative position of sun, earth and satellite. There is no
  fundamental obstacle to calculate and incorporate that pattern.
\item We assume a uniform initial charge between 70\% and 90\% of full capacity with identical bound and available charge.  Since the satellite needs to be switched off for transportation into space, assuming an equilibrated battery is valid. Being a single experiment, the GOMX-1 had a
  particular initial charge (though unknown). The charge of the
  orbiting battery can only be observed indirectly, by the voltage
  sustained.
\item We assume that the relative distance to a base station is a
  random quantity, and thus interpret several of the above statistics
  probabilistically. In reality, the position of the base station for
  GOMX-1 is at a particular fixed location (Aalborg, Denmark).
  Our approach can either be viewed as a kind of probabilistic
  abstraction of the relative satellite position and uncertainty of
  signal transmission, or it can be seen as reflecting that base
  stations are scattered around the planet. This especially would be a
  realistic in scenarios where satellite-to-satellite communication is
  used.
\item We assume that the satellite has no protection against battery
  depletion. In reality, the satellite has $2$ levels of software
  protection, activated at voltage levels $7.2$ and $6.5$,
  respectively, backed up by a hardware protection activated at $6$~V.
In these protection modes, various non-mission-critical functionality is switched-off. 
Despite omitting such power-saving modes, we still obtain conservative
guarantees on the probability that the battery powers the
satellite.
\end{itemize}

\paragraph*{Satellite model}

\begin{figure}[t]
  \centering
  \begin{tikzpicture}[->,>=stealth',shorten >=1pt,auto,node distance=2cm,
  thick,main node/.style={ellipse,fill=blue!20,draw,font=\sffamily\bfseries}]

  \node[main node, label=60:{190 mA},label=110:{$90$ min.}] (1) {Middle};
  \node[initial,main node, label=135:{90 mA},label=225:{$90$ min.}] (2) [left of=1] {Low};
  \node[main node, label=-45:{400 mA},label=-135:{$5$ min.}] (4) [below of=1] {Transfer};
  \node[main node, label=45:{250 mA},label=-45:{$90$ min.}] (3) [right of=1] {High};

  \path[every node/.style={font=\sffamily\small}]
  	(1) edge [bend left] node [left,pos=0.7] {$\frac{3}{5}$} (4)
  	(4) edge [bend left] node [right,pos=0.7] {$\frac{1}{8}$} (1)
    (1) edge [loop above] node {$\frac{2}{5}$} (1)
    (3) edge [bend left] node [left=3,pos=0.6] {$\frac{3}{5}$} (4)
    (3) edge [loop above] node {$\frac{2}{5}$} (3)
    (4.30) edge [bend left] node[right=3,pos=0.6] {$\frac{1}{4}$} (3)
    (2) edge [bend left] node [left,pos=0.8] {$\frac{3}{5}$} (4.150)
    (4) edge [bend left] node [auto,swap,pos=0.7] {$\frac{1}{8}$} (2)
    (2) edge [loop above] node {$\frac{2}{5}$} (2)
    (4) edge [loop right] node {$\frac{1}{2}$} (4);
\end{tikzpicture}
\caption{Markov task process of the load on the satellite. All load
  distributions are normal with mean depicted next to the states and
  with standard deviation $5$. This load is superposed with a strictly
  periodic load modelling charge by solar power infeed.}\label{fig:mc}
\end{figure}
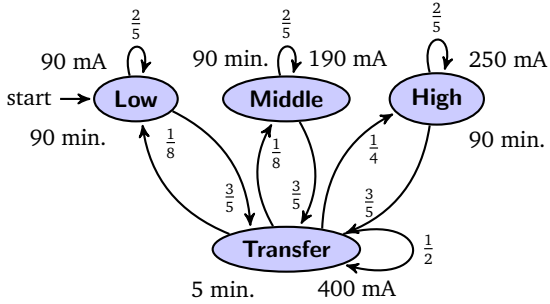

\begin{figure*}
\centering
\includegraphics[width=0.245\textwidth]{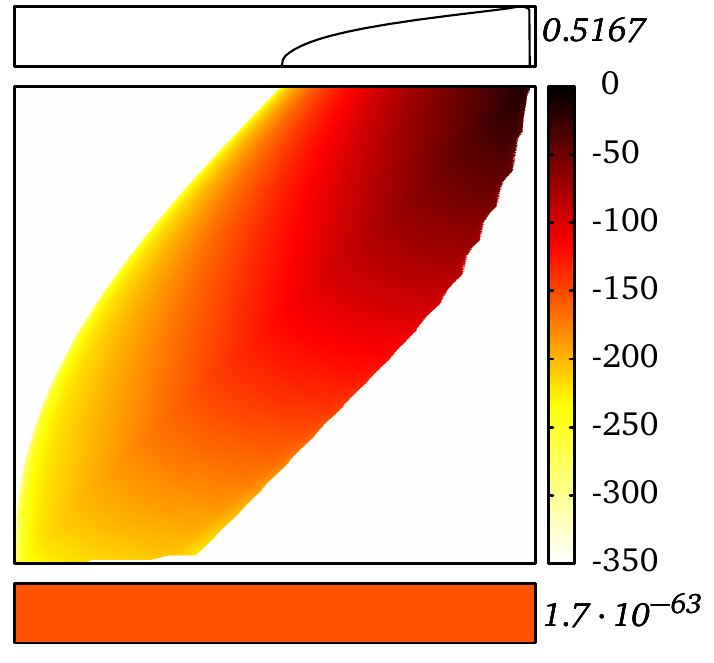}
\includegraphics[width=0.245\textwidth]{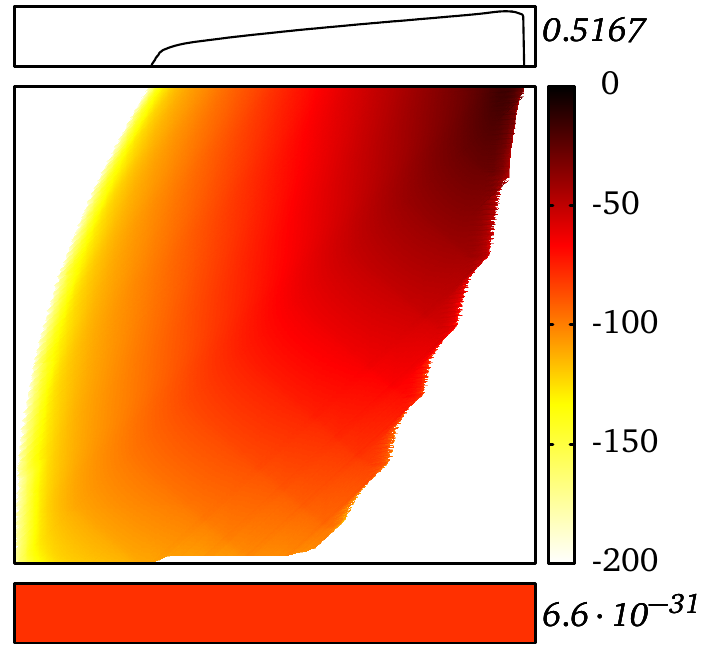}
\includegraphics[width=0.245\textwidth]{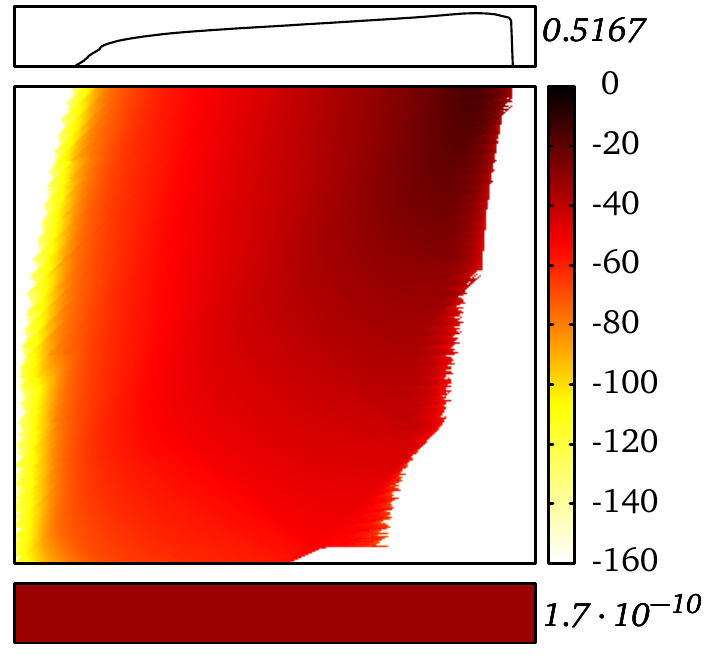}
\includegraphics[width=0.245\textwidth]{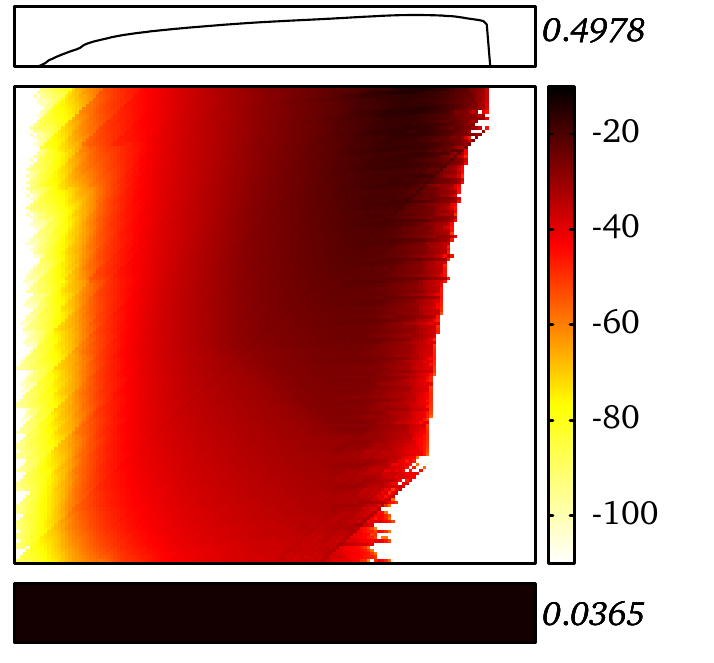}
\caption{\soc of the satellite after $1$ year \protect\footnotemark with different sizes of the battery. The leftmost \soc is with the original battery capacity, $5000$~mAh. In each further plot, the battery capacity is halved, i.e. $2500$~mAh, $1250$~mAh, and $625$~mAh.
Note that all the densities are depicted on the logarithmic scale (numbers in the legend stand for the order of magnitude). 
%
We observe that only the smallest battery does not give sufficient guarantees. Its probability of depletion after $1$ year is $0.0365$; the probability gets down to $1.7 \cdot 10^{-10}$ already for the $1250$~mAh battery.
The smaller the battery, the more crucial is the distinction of available and bound charge as a larger area of the plots is filled with non-trivial density.
%
%
}
\label{figBatterySize}
\end{figure*}
\footnotetext{Actually it is after 364 days, as this is in the middle of the charging phase. After 365 days the satellite is in eclipse and no density is exibited in the upper diagram.}

According to the above discussion, the load on the satellite is the
superposition of two piecewise constant loads.
\begin{itemize}
\item A probabilistic load reflecting the different operation modes, modeled by a Markov task process \mtp as depicted in Figure \ref{fig:mc}.%
\item A strictly periodic charge load alternating between $66$ minutes at $-400$ mA, and the remaining $33$ minutes at $0$ mA.%
\end{itemize}
One can easily express the charging load as another independent Markov
task process (where all probabilities are $1$) and consider the sum
load generated by these two processes in parallel (methods in
Section~\ref{sec:mtp} adapt straightforwardly to this
setting).

The \kibam in our model has following parameters:
\begin{itemize}
\item the ratio of the available charge $c = 1/2$ (artificially chosen value as parameters fitted by experiments on similar batteries strongly vary \cite{wognsen2014battery,so75079});
\item the diffusion rate $p = 0.0006$ per minute (we decreased the
  value reported by experiments \cite{so75079} by a factor of $4$
  because of the low average temperature in orbit, $3.5^{\circ}$C, and
  the influence of the Arrhenius equation \cite{liaw2003correlation}).
\end{itemize}

\paragraph*{Computational Aspects}

We implemented the continuous solution developed in the previous
sections in a high-level computational language Octave. This showed up
to be practical only up to sequences of a handful of tasks.
Therefore, we implemented a solution over a discretized abstraction of
the stochastic process induced by the MTP and the battery. By fixing
the number of discretization steps $K \in \Nats$ which yields the
discretization step $\delta = \frac{\capacity}{2}\cdot\frac{1}{K}$, we
obtain battery states
\begin{itemize}
\item $\{(n,m) \mid 0 < n < K,\; 0 <m < K\}$ in the inner space where each $(n,m)$ represents the adjacent rectangle of \emph{higher} charge $[n\delta,(n+1)\delta)\times[m\delta,(m+1)\delta)$,
\item states $\{(K,m) \mid 0 < m < K\}$ on the capacity boundary where each $(K,m)$ represents the adjacent line of \emph{higher} bound charge  $\{\frac{\capacity}{2}\} \times [m\delta, (m+1)\delta)$, and
\item state $(0,0)$ for the rest, $\{(a,b) \mid \text{$a < \delta$ or $b < \delta$}\}$.
\end{itemize}

We always represent higher charge by lower charge
(i.e. under-approximate \soc) since we are interested in
\emph{guarantees} on probabilities that the battery powers the MTP for
a given time horizon.
Similarly, we replace load distributions by discrete distributions where each point represents an adjacent left interval (i.e., we over-approximate the load).
The continuous methods of Lemma~\ref{lem:random-kibam-with-bounds} are
 easily adapted to this discrete setting, basically replacing integrals by finite sums. 

 This methods gives us an underapproximation of the probability that
 the battery powers the satellite. We do not have any prior error
 bound, but one can make the results arbitrarily precise by increasing
 $K$, at the price of quadratic cost increase.

Our implementation is done in \textsc{C++}, we used $K =$ 1200, 600, 300 and 150 for the experiments with the batteries of capacity $5000$ mAh, $2500$ mAh, $1250$ mAh and $625$ mAh, respectively to guarantee equal relative precision.
All the experiments have been performed on a machine equipped with an Intel Core i5-2520M CPU @ 2.50GHz and 4GB RAM. All values occuring are represented and calculated with standard IEEE 754 double-precision binary floating-point format except for the values related to the battery being depleted where we use arbitrary precision arithmetic (as to this number, we keep adding values from the inner area that are of much lower order of magnitude).

%
\paragraph*{Model evaluation}

\begin{figure}
\centering
\includegraphics[width=0.225\textwidth]{1250.png}
\hspace{0.5em}
\includegraphics[width=0.225\textwidth]{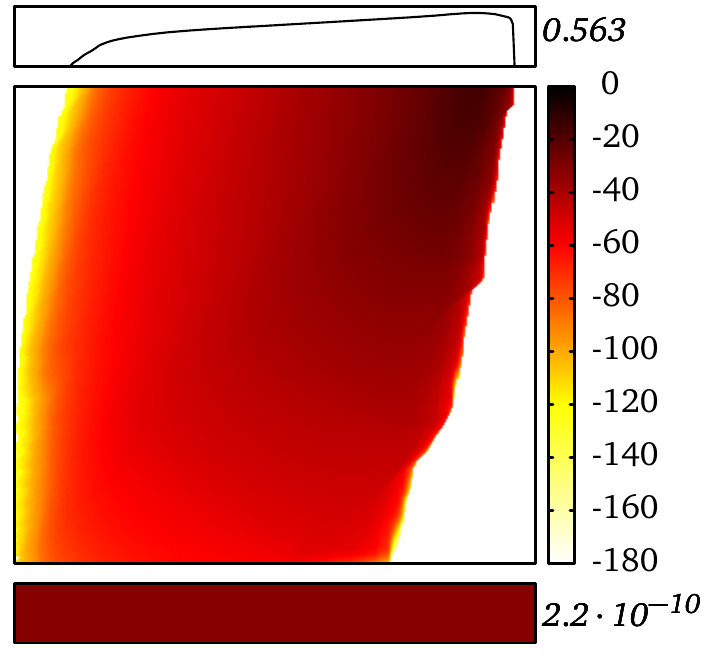}
\caption{Load noise. The $1$ year run of the $1250$~mAh battery with Dirac loads on the left and with noisy loads on the right. We used Gaussian noise with standard deviation $5$.}
\label{figBatteryNoise}
\end{figure}

\begin{figure}
\begin{center}
\includegraphics[width=0.225\textwidth]{5000.png}
\hspace{0.5em}
\includegraphics[width=0.225\textwidth]{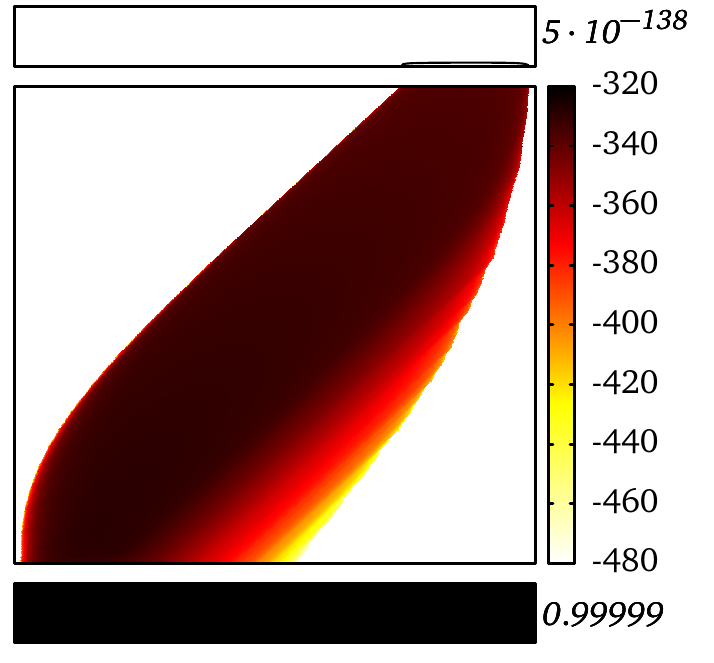}
\end{center}
\caption{Number of solar panels. The full $5000$~mAh battery with $9$
  solar panels on the left and $6$ solar panels on the right. This
  shows that for the current load, a $1$ unit cube design with solar
  panels on $6$ or less sides is not possible.}
\label{figBatterySolar}
\end{figure}

We performed various experiments with this model, to explore the
random \kibam technology. We here report on four distinct evaluations,
demonstrating that valuable insight into the model can be obtained.

\begin{enumerate}
\item The $5000$~mAh battery in the real satellite is known to be
  over-dimensioned. Our aim was to find out how much. Hence, we
  performed a sequence of experiments, decreasing the size of the
  battery exponentially. The results are displayed and explained in
  Figure~\ref{figBatterySize}. We found out that $1/4$ of the
  capacity still provides sufficient guarantees to power the satellite
  for 1 year while $1/8$ of the capacity, $625$ mAh, does not. 
\item We compared our results with a simple linear battery model of
  the same capacity. This linear model is not uncommon in the
  satellite domain, it has for instance been used in the
  \emph{Envisat} and \emph{CryoSat} missions~\cite{gilles}. We obtain
  the following probabilities for  battery depletion:
   \begin{center}
   \begin{tabular}{c|cc}
\textbf{capacity}  & \textbf{linear battery model} & \textbf{\kibam} \\
  \hline\hline 5000 mAh \rule{0pt}{2.6ex}& $1.86 \cdot 10^{-84}$ & $1.7\cdot 10^{-63}$ \\ 
        625 mAh & $2.94\cdot 10^{-8}$ & $0.0365$ \\ 
  \end{tabular}
  \end{center}
  The linear model turns out to be surprisingly (and likely
  unjustifiably) optimistic, especially for the $625$ mAh battery.
\item We (computationally) simplified the two experiments above by assuming Dirac loads.
%
To analyze the effect of the white noise, we compared the Dirac loads with the noisy loads, explained earlier, on the $625$ mAh battery.
As expected, the noise (a) smoothes out the distribution a little and (b) pushes a bit more of the distribution to full and empty states, see Figure \ref{figBatteryNoise}.
\item Our reference satellite is a two-unit satellite, i.e. is built from two cubes, each $10$ cm per side. In the current design, $9$ of the $10$ external sides are covered by solar panels, the remaining one is used for both radio antenna and camera.
We thus analyzed whether a one-unit design with only $5$ solar panels is possible.
The answer is negative, the system runs out of energy rapidly with high probability.
Figure \ref{figBatterySolar} displays that even for $6$ panels the charge level is highly insufficient to sustain the  load.
\end{enumerate}

\section{Alternative Approaches}
The results reported above are obtained from a discretized
abstraction of the stochastic process induced by the MTP and the
battery, solved numerically and with high-precision arithmetic where
needed. 

One could instead consider estimating the probability $\dead{t}$ of
the battery depletion using ordinary simulation
techniques~\cite{gillespie1976general}. Considering a
battery of capacity 5000 mAh, this would mean that about $10^{63}$
simulations traces are needed on average to observe the rare event of
a depleted battery at least once. This seems prohibitive, also if
resorting to massively parallel simulation, which may reduce the
exponent by a small constant at most. A possible way out of this might
lie in the use of rare event simulation techniques to speed up
simulation~\cite{villen1994restart}.

The behaviour of \kibam with capacity bounds can be expressed as a
relatively simple \emph{hybrid automaton}
model~\cite{henzinger2000theory}. Similarly, the random \kibam with
capacity bounds can be regarded as an instance of a \emph{stochastic
  hybrid system} (SHS)
\cite{abate2008probabilistic,altman1997asymptotic,blom2006stochastic,bujorianu2005bisimulation,davis1984piecewise,sproston2000decidable}. This
observation opens some further evaluation avenues, since there are
multiple tools available
publicly for checking reachability properties of SHS. In particular,
\textsc{Faust}$^2$~\cite{FAUST13}, \textsc{SiSat}~\cite{FraHerTei08:QAPL:SSMT} and
\textsc{ProHVer}~\cite{ZhangSRHH10, FraenzleHHWZ11} appear adequate at first sight. Our experiments
with \textsc{Faust}$^2$ however were unsuccessful, basically due to a model
mismatch: The tool thus far assumes stochasticity in all dimensions,
because it operates on stochastic kernels, while our model is
non-stochastic in the bound charge dimension. With \textsc{SiSat}, we so far failed to
encode the MTP (or its effect) into an input accepted by the tool. The
MTP can be considered as a compact description of an otherwise
intricate semi-Markov process running on a discrete time line. This is
in principle supported by \textsc{SiSat}, yet we effectively failed to provide
a compact encoding. Our \textsc{ProHVer} experiments failed for a different
reason, namely the sheer size of the problem.  All the above tools
have not been optimized for dealing with very low probabilities as
they appear in the satellite case.

\section{Conclusion}

Inspired by the needs of an earth-orbiting satellite mission, we
extended in this paper the theory of kinetic battery models in two
independent dimensions. First, we addressed battery charging up to
full capacity. Second, we extended the theory of the \kibam
differential equations to a stochastic setting. We provided a symbolic
solution for random initial \soc and a sequence of
piecewise-constant random loads.

These sequences can be generated by a stochastic process representing
an abstract and averaged behavioral model of a nano satellite operating in earth orbit,
superposed with a deterministic representation of the solar infeed in
orbit. We illustrated the approach by several experiments performed on
the model, especially varying the size of the battery, but also the
number of solar panels. 

%
ESA is running a large educational program~\cite{esawebsite} for
launching missions akin to GOMX-1.
The satellites are designed by student teams, have the form of standardized 1 unit cube with maximum mass of 1 kg, and target mission times of up to four years. 
The random \kibam
presented here is of obvious high relevance for any participating
team. It can help quantify the risk of premature depletion for the
various battery dimensions at hand, and thereby enable an optimal use
of the available weight and space budget. Our experiments show that
using the simpler linear battery model instead is far too optimistic
in this respect.

For a fixed setup, one can also use the technology offered by us for
optimal task scheduling: In the same way as we can follow a single
\soc distribution, we can also branch into several distributions and
determine which of them is best according to some metric. Taking
inspiration from~\cite{wognsen2014battery}, this can be
combined with statistical model checking so as to find the optimal
task schedule of a given set of tasks.

Several extensions can and should be integrated in the model. Among
them, temperature dependencies are of particular interest. 
A temperature change has namely opposing physical effects
in solar panels and in the battery, having intriguing consequences such as piecewise exponential decay in the charging process.
%
An extension that is particularly important for long lasting missions, is incorporating a model of {battery wearout}.
%
So far we assume the battery capacity to be constant
along the mission time. Notably, our contribution is the first to
consider capacity bounds in operation at all, as far as we are aware.

\paragraph*{Acknowledgements} The authors are grateful for inspiring
discussions with Peter Bak (GomSpace ApS), Erik R.\ Wognsen (Aalborg
University), and other members of the SENSATION consortium, as well as
with Pascal Gilles (ESA Centre for Earth Observation), Xavier Bossoreille (Deutsches Zentrum f\"ur Luft- und Raumfahrt%
) and Marc Bouissou (\'Electricit\'e de France S.A.%
, \'Ecole Centrale Paris - LGI). This work is
supported by the Transregional Collaborative Research Centre SFB/TR 14
AVACS, and the 7th EU Framework Program under grant agreements
295261 (MEALS) and 318490 (SENSATION).

\bibliographystyle{abbrv}
\bibliography{main}

\begin{thebibliography}{10}

\bibitem{abate2008probabilistic}
A.~Abate, M.~Prandini, J.~Lygeros, and S.~Sastry.
\newblock Probabilistic reachability and safety for controlled discrete time
  stochastic hybrid systems.
\newblock {\em Automatica}, 44(11):2724--2734, 2008.

\bibitem{altman1997asymptotic}
E.~Altman and V.~Gaitsgory.
\newblock Asymptotic optimization of a nonlinear hybrid system governed by a
  markov decision process.
\newblock {\em SIAM Journal on Control and Optimization}, 35(6):2070--2085,
  1997.

\bibitem{blom2006stochastic}
H.~A. Blom, J.~Lygeros, M.~Everdij, S.~Loizou, and K.~Kyriakopoulos.
\newblock {\em Stochastic hybrid systems: theory and safety critical
  applications}, volume 337.
\newblock Springer Heidelberg, 2006.

\bibitem{boker2014battery}
U.~Boker, T.~A. Henzinger, and A.~Radhakrishna.
\newblock Battery transition systems.
\newblock In {\em Proceedings of the 41st annual ACM SIGPLAN-SIGACT symposium
  on Principles of programming languages}, pages 595--606. ACM, 2014.

\bibitem{buchmann}
I.~Buchmann.
\newblock {\em Batteries in a portable world}.
\newblock Cadex Electronics Richmond, 2001.

\bibitem{bujorianu2005bisimulation}
M.~L. Bujorianu, J.~Lygeros, and M.~C. Bujorianu.
\newblock Bisimulation for general stochastic hybrid systems.
\newblock In {\em Hybrid Systems: Computation and Control}, pages 198--214.
  Springer, 2005.

\bibitem{DBLP:conf/dsn/ClothJH07}
L.~Cloth, M.~R. Jongerden, and B.~R. Haverkort.
\newblock Computing battery lifetime distributions.
\newblock In {\em The 37th Annual {IEEE/IFIP} International Conference on
  Dependable Systems and Networks, {DSN} 2007, 25-28 June 2007, Edinburgh, UK,
  Proceedings}, pages 780--789. {IEEE} Computer Society, 2007.

\bibitem{DBLP:journals/adcm/CorlessGHJK96}
R.~M. Corless, G.~H. Gonnet, D.~E.~G. Hare, D.~J. Jeffrey, and D.~E. Knuth.
\newblock On the lambert\emph{W} function.
\newblock {\em Adv. Comput. Math.}, 5(1):329--359, 1996.

\bibitem{davis1984piecewise}
M.~H. Davis.
\newblock Piecewise-deterministic markov processes: A general class of
  non-diffusion stochastic models.
\newblock {\em Journal of the Royal Statistical Society. Series B
  (Methodological)}, pages 353--388, 1984.

\bibitem{esawebsite}
Esa.
\newblock Esa cubesat program, Oct. 2014.
\newblock \url{http://www.esa.int/Education/CubeSats}.

\bibitem{FAUST13}
S.~{Esmaeil Zadeh Soudjani}, C.~Gevaerts, and A.~Abate.
\newblock Faust2: Formal abstractions of uncountable-state stochastic
  processes.
\newblock {\em arXiv preprint}, 2014.
\newblock arXiv:1403.3286.

\bibitem{DBLP:conf/ijcai/FoxLM11}
M.~Fox, D.~Long, and D.~Magazzeni.
\newblock Automatic construction of efficient multiple battery usage policies.
\newblock In T.~Walsh, editor, {\em {IJCAI} 2011, Proceedings of the 22nd
  International Joint Conference on Artificial Intelligence, Barcelona,
  Catalonia, Spain, July 16-22, 2011}, pages 2620--2625. {IJCAI/AAAI}, 2011.

\bibitem{FraenzleHHWZ11}
M.~Fr{\"a}nzle, E.~M. Hahn, H.~Hermanns, N.~Wolovick, and L.~Zhang.
\newblock Measurability and safety verification for stochastic hybrid systems.
\newblock In {\em HSCC}, pages 43--52, New York, NY, USA, 2011. ACM Press.

\bibitem{FraHerTei08:QAPL:SSMT}
M.~Fr{\"a}nzle, H.~Hermanns, and T.~Teige.
\newblock Stochastic satisfiability modulo theory: A novel technique for the
  analysis of probabilistic hybrid systems.
\newblock In M.~Egerstedt and B.~Mishra, editors, {\em Pre-Proceedings of the
  European Joint Conferences on Theory and Practice of Software (ETAPS) 2008
  Sixth Workshop on Quantitative Aspects of Programming Languages (QAPL 2008)},
  volume 4981 of {\em Lecture Notes in Computer Science (LNCS)}, pages
  172--186. Springer-Verlag, 2008.
\newblock Extended abstract.

\bibitem{gilles}
P.~Gilles.
\newblock Private communication.
\newblock 2014.

\bibitem{gillespie1976general}
D.~T. Gillespie.
\newblock A general method for numerically simulating the stochastic time
  evolution of coupled chemical reactions.
\newblock {\em Journal of computational physics}, 22(4):403--434, 1976.

\bibitem{gomspacewebsite}
GomSpace.
\newblock Gomspace gomx-1, Oct. 2014.
\newblock \url{http://gomspace.com/?p=gomx1}.

\bibitem{henzinger2000theory}
T.~A. Henzinger.
\newblock {\em The theory of hybrid automata}.
\newblock Springer, 2000.

\bibitem{jongerden2009maximizing}
M.~Jongerden, B.~Haverkort, H.~Bohnenkamp, and J.~Katoen.
\newblock Maximizing system lifetime by battery scheduling.
\newblock In {\em Dependable Systems \& Networks, 2009. DSN'09. IEEE/IFIP
  International Conference on}, pages 63--72. IEEE, 2009.

\bibitem{so75079}
M.~R. {Jongerden}.
\newblock {\em Model-based energy analysis of battery powered systems}.
\newblock PhD thesis, Enschede, December 2010.

\bibitem{DBLP:journals/iee/JongerdenH09}
M.~R. Jongerden and B.~R. Haverkort.
\newblock Which battery model to use?
\newblock {\em {IET} Software}, 3(6):445--457, 2009.

\bibitem{lee1987synchronous}
E.~A. Lee and D.~G. Messerschmitt.
\newblock Synchronous data flow.
\newblock {\em Proceedings of the IEEE}, 75(9):1235--1245, 1987.

\bibitem{liaw2003correlation}
B.~Y. Liaw, E.~P. Roth, R.~G. Jungst, G.~Nagasubramanian, H.~L. Case, and D.~H.
  Doughty.
\newblock Correlation of arrhenius behaviors in power and capacity fades with
  cell impedance and heat generation in cylindrical lithium-ion cells.
\newblock {\em Journal of power sources}, 119:874--886, 2003.

\bibitem{manwell1993lead}
J.~F. Manwell and J.~G. McGowan.
\newblock Lead acid battery storage model for hybrid energy systems.
\newblock {\em Solar Energy}, 50(5):399--405, 1993.

\bibitem{rao}
V.~Rao, G.~Singhal, A.~Kumar, and N.~Navet.
\newblock Battery model for embedded systems.
\newblock In {\em 18th International Conference on {VLSI} Design {(VLSI} Design
  2005), with the 4th International Conference on Embedded Systems Design, 3-7
  January 2005, Kolkata, India}, pages 105--110. {IEEE} Computer Society, 2005.

\bibitem{PhysRevLett.81.4660}
J.~Rodr{\'\i}guez-Carvajal, G.~Rousse, C.~Masquelier, and M.~Hervieu.
\newblock Electronic crystallization in a lithium battery material: Columnar
  ordering of electrons and holes in the spinel ${\mathrm{limn}}_{2}{O}_{4}$.
\newblock {\em Phys. Rev. Lett.}, 81:4660--4663, Nov 1998.

\bibitem{sahner1987performance}
R.~A. Sahner and K.~S. Trivedi.
\newblock Performance and reliability analysis using directed acyclic graphs.
\newblock {\em IEEE Transactions on Software Engineering}, 13(10):1105--1114,
  1987.

\bibitem{sproston2000decidable}
J.~Sproston.
\newblock Decidable model checking of probabilistic hybrid automata.
\newblock In {\em Formal Techniques in Real-Time and Fault-Tolerant Systems},
  pages 31--45. Springer, 2000.

\bibitem{theelen2006scenario}
B.~D. Theelen, M.~Geilen, T.~Basten, J.~P. Voeten, S.~V. Gheorghita, and
  S.~Stuijk.
\newblock A scenario-aware data flow model for combined long-run average and
  worst-case performance analysis.
\newblock In {\em Formal Methods and Models for Co-Design, 2006. MEMOCODE'06.
  Proceedings. Fourth ACM and IEEE International Conference on}, pages
  185--194. IEEE, 2006.

\bibitem{villen1994restart}
M.~Vill{\'e}n-Altamirano and J.~Vill{\'e}n-Altamirano.
\newblock Restart: a straightforward method for fast simulation of rare events.
\newblock In {\em Simulation Conference Proceedings, 1994. Winter}, pages
  282--289. IEEE, 1994.

\bibitem{wognsen2014battery}
E.~R. Wognsen, R.~R. Hansen, and K.~G. Larsen.
\newblock Battery-aware scheduling of mixed criticality systems.
\newblock In {\em Leveraging Applications of Formal Methods, Verification and
  Validation. Specialized Techniques and Applications}, pages 208--222.
  Springer Berlin Heidelberg, 2014.

\bibitem{ZhangSRHH10}
L.~Zhang, Z.~She, S.~Ratschan, H.~Hermanns, and E.~M. Hahn.
\newblock Safety verification for probabilistic hybrid systems.
\newblock In {\em CAV}, volume 6174 of {\em LNCS}, pages 196--211. Springer,
  2010.

\end{thebibliography}


\end{document}